\documentclass[%
 reprint,
 amsmath,amssymb,
 aps,
]{revtex4-1}

\usepackage{graphicx}
\usepackage{bm}
\usepackage[protrusion=true,expansion=true]{microtype} 
\usepackage{wrapfig} 

\usepackage[T1]{fontenc} 
\usepackage{bm}

\usepackage{mathrsfs}
\usepackage{braket}
\usepackage{amsmath}
\usepackage{amssymb}
\usepackage{color}
\usepackage{indentfirst}
\usepackage{amsfonts}
\usepackage{nicefrac}

\newcommand{\be}{\begin{equation}}
\newcommand{\ee}{\end{equation}}
\newcommand{\bea}{\begin{equation}\begin{aligned}}
\newcommand{\eea}{\end{aligned}\end{equation}}
\newcommand{\beS}{\begin{equation}\begin{split}}
\newcommand{\eeS}{\end{split}}
\newcommand{\bV}{\begin{pmatrix}}
\newcommand{\eV}{\end{pmatrix}}
 \newcommand{\norm}[1]{\left\lVert#1\right\rVert}

\DeclareMathOperator{\tr}{tr}

\newcommand{\de}{\partial}
\newcommand{\Id}{\mathbb{I}}

\newcommand{\Li}{\mathscr{L}}

\newcommand{\D}{\mathcal{D}}

\newcommand{\p}{_{\uparrow}}
\newcommand{\m}{_{\downarrow}}

\newcommand{\hot}{^{hot}}
\newcommand{\cold}{^{cold}}
\newcommand{\f}{\text{f}}
\newcommand{\z}{\text{z}}
\newcommand{\dd}{\text{d}}

\newcommand{\T}{\mathcal{T}}

\begin{document}

\title{Exact solution of time-dependent Lindblad equations with closed algebras}

\author{Stefano Scopa$^{1}$}
\email{scopa1@univ-lorraine.fr} 

\author{Gabriel T. Landi$^{2}$}
\email{gtlandi@if.usp.br} 

\author{Adam Hammoumi$^{1}$}

\author{Dragi Karevski $^{1}$}
\email{dragi.karevski@univ-lorraine.fr}

 \affiliation{$^{1}$ Laboratoire de Physique et Chimie Th\'eoriques, CNRS, UMR No. 7019, Universit\'e de Lorraine, BP 239, 54506 Vandoeuvre-l\'es-Nancy Cedex, France}
\affiliation{$^{2}$ Instituto de F\'isica, Universidade de S\~ao Paulo, 05314-970 S\~ao Paulo, S\~ao Paulo, Brazil}

\date{\today}

\begin{abstract}
Time-dependent Lindblad master equations have important applications in areas ranging from quantum thermodynamics to dissipative quantum computing. In this paper we outline a general method for writing down exact solutions of time-dependent Lindblad equations whose superoperators form closed algebras. We focus on the particular case of a single qubit and study the exact solution generated by both coherent and incoherent mechanisms. We also show that if the time-dependence is periodic, the problem may be recast in terms of Floquet theory. As an application, we give an exact solution for a two-levels quantum heat engine operating in a finite-time.
\end{abstract}

\maketitle


\section{Introduction}

%
%
%
%
%

Understanding the non-equilibrium aspects emerging during the time evolution of driven quantum systems is one of the main questions that the theoretical physics community have been trying to address in the last decades.
This large interest is also motivated by the enormous improvements in the experimental manipulation of quantum platforms such as cold atoms \cite{coldatoms1,coldatoms2,coldatoms3,coldatoms4,coldatoms5,coldatoms6,coldatoms7,coldatoms8,coldatoms9,coldatoms10,coldatoms11,coldatoms12,coldatoms13,coldatoms14} or quantum spin chains \cite{spin_chainEXP1,spin_chainEXP2,spin_chainEXP3,spin_chainEXP4,spin_chainEXP5,spin_chainEXP6,spin_chainEXP7}. 
Much of the effort, however, has so far focused on  driven closed quantum systems (c.f. the reviews~\cite{rev-off-eq1,rev-off-eq2} and references quoted therein).
Studies on open quantum systems still remain scarce \cite{Thingna2017,Baker2018,Kossloff2018,per1,per2,per3,per4,per5,per6,per7,per8,per9,Prosen2018,Dai-16,Dai-17,Scopa-18,Alicki2012,Alicki2006,
Rau-2002,Rau-2003,Rau-2005,Haddadfarshi-15,Schnell18}. 
%
%
%
%
%
%
The reason is that modelling a quantum system which exchanges informations with an environment  runs into difficulties due to the large number of degrees of freedom in question. 
%
%
%

An alternative approach is to consider only the effective role of the environment described in terms of a complete positive trace-preserving (CPTP) map acting on the system. 
This abstraction turns out to be particularly fruitful, specially for Markovian systems in which the evolution can be cast in Lindblad form \cite{Lindblad-75,Lindblad-76, GKS-76,Breuer-Petruccione,Alicki-Lendi}. 
In this sense, by allowing the Lindblad generators to be time-dependent, one can model phenomenologically the evolution of Markovian open quantum systems driven by coherent and incoherent mechanisms. 
In addition to their general theoretical interest, time-dependent Lindblad equations have also proved useful as tools in areas such as quantum optics, quantum information \cite{quantum_info} and quantum thermodynamics \cite{Anders17,Kosloff_rev,AlickiKossloff18}. 

In the last years, different techniques have been developed to solve these master equations, such as exact diagonalization \cite{exact_diag1,exact_diag2,exact_diag3}, algebraic methods \cite{algebraic_method1,algebraic_method2,algebraic_method3}, Bloch-like equations \cite{Rau-2002,Rau-2003,Rau-2005}, series expansions \cite{dyson_series_exp1,dyson_series_exp2} and Floquet theory for periodic drivings. 
%
%
%
%
The case of Floquet dynamics of periodic systems deserves special attention. 
Naturally motivated by the enormous success of Floquet theory \cite{review_Floq} in closed systems  \cite{Rahav03,Goldman-PRX-14,Uper1,Uper2,Uper3,Uper4,Uper5,Uper6,Uper7,Uper8,Uper9,Uper10,Uper11,SUK18},  there has recently been  substantial effort in extending it to open dynamics \cite{per1,per2,per3,per4,per5,per6,per7,per8,per9,Prosen2018,Dai-16,Dai-17}. 
Unfortunately, in the case of open dynamics, one may run into mathematical difficulties concerning the existence of a unique generator for the stroboscopic dynamics. This was beautifully illustrated in Refs.~\cite{per1,Schnell18} and also discussed in Ref.~\cite{Haddadfarshi-15}, where the authors propose a method to generate high frequency Magnus expansions which avoid these types of problems.

%


%


With the goal of shedding light on this difficult problem, in  this paper we propose a general method for the solution of time-dependent Lindblad equations based on the closure of the superoperator's algebra. We first prove that a markovian open quantum system of finite Hilbert space dimension always has a set of time-independent superoperators which form a closed algebra. Afterwards, we show how to move to a generalised rotating frame where the Lindblad generator is time-independent and the dynamics can be easily solved. We also discuss the specific case of a periodic driving where our framework coincides with  Lindblad-Floquet theory. 
As our main application, we apply our results  to describe the operation of a quantum heat engine operating at finite time \cite{Alicki-79,Kosloff-Levy,Alicki2015,Rezek2006}. 

The paper is organised as follows. In the next section, we introduce the time-dependent Lindblad equation and we construct for it a closed algebra of superoperators, key object of our method. Then, we present the framework and its application to a periodic driving, see Sec.\ref{periodic_driving}. Here, we show how it is possible to cast the initial problem in the form of a Floquet theory. In Sec.\ref{2lvl}, we propose an application to the case of a single qubit with two explicit pedagogical examples. We move to finite-time quantum heat-engines in Sec.\ref{quantum_heat_engines}. After discussing the improvements carried by our approach, we solve in details the operation of a 2-levels heat-engine, see Sec.\ref{2lvl_engine}, focusing on the case of a Carnot and of an Otto cycle. A summary of this work with some concluding remarks is given in Sec.\ref{conclusions}. We leave some technical aspects and the comparison with other known techniques to the appendices.

\section{The method}
We investigate the dynamics of an open quantum system ${\cal S}$ of finite Hilbert space dimension $\text{dim}\left(\mathscr{H}_{\cal S}\right)=n<\infty$ in the case in which both Hamiltonian and non-Hamiltonian parameters have a generic time dependence. In the markovian approximation, the time evolution of the reduced density matrix $\rho_t$ can be written in terms of a master equation of Lindblad form \cite{Lindblad-75,Lindblad-76, GKS-76}
\be\label{Lind_eq}
\frac{d}{dt}\rho_t =-i[H_t,\rho_t] + \sum_{j=1}^{n^2-1} \Gamma_j(t) ( L_j \rho_t L_j^{\dagger} - \frac{1}{2} \{ L^{\dagger}_j L_j, \rho_t \} ),
\ee
where $H_t$ is the effective time-dependent Hamiltonian of the open system and $L_j \in \text{End}(\mathscr{H}_{\cal S})$ are jump operators accounting for relevant elementary dissipative processes having a relaxation time $\sim \Gamma_j^{-1}(t)$, with $\Gamma_j\geq 0$ $\forall \, t\geq t_0$. In our notations, $\{a,b\}\equiv ab -ba$ is the anticommutator of operators $a,b \in \text{End}(\mathscr{H}_{\cal S})$.
Notice that the non-Hamiltonian generator in Eq. $\eqref{Lind_eq}$ is a quadratic form and thus it remains invariant under a change of basis in $\text{End}(\mathscr{H}_{\cal S})$.  Therefore, we consider a complete set of operators $\left\{ F_k \right\}_{k=0}^{n^2-1}$ of $\text{End}\left(\mathscr{H}_S\right)$ satisfying:
\be
 \tr(F_k)=\delta_{k,0}\,, \qquad   \tr(F_k^{\dagger} \,F_l)=\delta_{kl} 
\ee
where $F_0= \Id_n /n$ is proportional to the identity operator $\Id_n$ and $F_{k}\equiv F_{k}^{\dagger}$ for $k>0$ are chosen to be the generators of the $\mathfrak{su}(n)$ Lie algebra. A general construction of the complete orthonormal set  of $\left\{ F_k \right\}_{k=0}^{n^2-1}$ is shown in  appendix \ref{app_complete_set}. In terms of this basis, the master equation $\eqref{Lind_eq}$ becomes
\be\begin{split}\label{Lind_eq_2}
\frac{d}{dt}\rho_t = &-i\sum_{j=1}^{n^2-1} h_j(t) [F_j, \rho_t] \\
&+ \sum_{k,l=1}^{n^2-1} \gamma_{kl}(t) ( F_k \rho_t F_l - \frac{1}{2} \{ F_l F_k , \rho_t \})
\end{split}\ee
where $h_j(t)=\tr\left(H_t \, F_j \right)$ and $\gamma(t)$ is a semipositive definite hermitian matrix $\forall \,t\geq t_0$. At this point, we introduce a vectorisation procedure that maps $\text{End}(\mathscr{H}_{\cal S})$ into an isomorphic and isometric vector space $\mathfrak{K}\sim \mathbb{C}^{n^2}$ such that
\be
\ket{\psi}\bra{\phi} \, \mapsto \ket{\phi} \otimes \ket{\psi}\,, \quad \forall\,\psi\,, \phi \in \mathscr{H}_{\cal S}\,.
\ee

Vectorising the master equation~$\eqref{Lind_eq}$ then yields
\be\label{Lind_vec}
\frac{d}{dt} \, \ket{\rho_t} = \Li_t \, \ket{\rho_t}
\ee
with Liouvillean generator $\Li_t$ 
\be\label{Li}
\Li_t= \sum_{j=1}^{n^2-1} h_j(t) \, {\cal H}_j + \sum_{k,l=1}^{n^2-1} \gamma_{kl}(t) \,{\cal D}_{kl}
\ee
where
\be\label{sup_basis1}
{\cal H}_j=-i (\Id_n \otimes F_j - F_j^* \otimes \Id_n)
\ee
and
\be\label{sup_basis2}
{\cal D}_{kl}=F_l^* \otimes F_k -\frac{1}{2} \Id_n \otimes F_l F_k -\frac{1}{2} F_k^*F_l^* \otimes \Id_n \,.
\ee
The set of superoperators  $\{ \{{\cal H}_j\}_{j=1}^{n^2-1},\, \{ {\cal D}_{kl}\}_{k,l=1}^{n^2-1}\}$ forms a basis for a generic Liouvillean $\Li_t$ and has a closed Lie algebraic structure.
The general form of this algebra is somewhat complicated and is reported in appendix \ref{algebra_superop}.
However, as we will show below when we discuss specific examples, it can be greatly simplified in practice.

Let us denote the dynamical map which solves Eq.~(\ref{Lind_vec})  as
\be
\ket{\rho_t}= \Lambda_{t,t_0} \, \ket{\rho_{t_0}}\,,
\ee
The fact that the algebra of the superoperators is closed  allows us to parametrise the map $\Lambda_{t,t_0}$
as an element of the Lie group associated to the algebra of the generators
\be\label{param_dyn_map}
\Lambda_{t,t_0}= \prod_{j=1}^{n^2-1} e^{\lambda_j(t) \, {\cal H}_j} \, \prod_{k,l=1}^{n^2-1} e^{\pi_{kl}(t) \, {\cal D}_{kl}}
\ee
with canonical coordinates $\lambda_j(t),\, \pi_{kl}(t)$ of the second kind. 
The main point now, is that the time-dependent solution is shifted entirely to the complex functions $\lambda_j(t)$ and $\pi_{kl}(t)$. 
Moreover, these parameters will satisfy a system of coupled differential equations. 
To see this more specifically, we may use Eq.~$\eqref{Lind_vec}$ to write
\be\label{non_unitary_int}
(\frac{d}{dt} \Lambda_{t,t_0})(\Lambda_{t,t_0})^{-1}= \Li_t
\ee
Using Baker-Campbell-Hausdorff (BCH) relations \cite{Baker, Campbell1,Campbell2, Hausdorff} (see also \cite{Hall-book}) on the l.h.s. together with the linear independence of the superoperators,  will then yield  a set of coupled first-order differential equations for the coordinates $\lambda_j(t),\, \pi_{kl}(t)$. 
Unfortunately, the general form of these equations is too complex for arbitrary algebras, so that we shall illustrate this only by looking at specific examples below.


\subsection{Rotating frame transformation}\label{rotating_frame}

Here we introduce a complementary approach to the solution of the  time-dependent Lindblad equation $\eqref{Lind_vec}$ based on the idea of a generalized rotating frame. Consider a time-dependent invertible map  $W_t$ that connects our open system ${\cal S}$ to an auxiliary frame $\widetilde{\cal S}$ :
\be
\ket{\rho_t} \, \mapsto \, \ket{\widetilde{\rho}_t}= W_t \ket{\rho}_t\,.
\ee
The evolution in $\widetilde{\cal S}$ is governed by 
\be
\frac{d}{dt}\, \ket{\widetilde{\rho}_t} = \widetilde{\Li}_t \ket{\widetilde{\rho}_t}
\ee
where 
\be\label{aux_Li}
\widetilde{\Li}= W_t \, \Li_t \, W_t^{-1} + (\frac{d}{dt} W_t)\, W_t^{-1}\,.
\ee
If we can design $W_t$ in such a way that $\widetilde{\Li}$ is time independent, then the dynamics in $\widetilde{\cal S}$ will simply be given by 
\be
\ket{\widetilde{\rho}_t}= e^{\widetilde{\Li}(t-t_0)} \ket{\widetilde{\rho}_{t_0}}
\ee
and moving back to the original frame ${\cal S}$ we conclude that
\be\label{rot_frame}
\Lambda_{t,t_0}= W_t^{-1} \,  e^{\widetilde{\Li}(t-t_0)} \, W_{t_0}\,.
\ee
Chosing $W_t$ in the form
\be\label{W_map}
W_t= \prod_{j=1}^{n^2-1} e^{f_j(t)\, {\cal H}_j} \prod_{k,l=1}^{n^2-1} e^{g_{kl}(t)\, {\cal D}_{kl}}\,,
\ee
with coordinates $f_j(t)$, $g_{kl}(t)$ and using BCH relations in the r.h.s. of Eq.$\eqref{aux_Li}$, $\widetilde{\Li}$ admits a closed expression
\be\label{aux_Li2}
\widetilde{\Li}= \sum_{j=1}^{n^2-1} \widetilde{h}_j \, {\cal H}_j + \sum_{k,l=1}^{n^2-1} \widetilde{\gamma}_{kl} \, {\cal D}_{kl}
\ee
in the basis of superoperators $\eqref{sup_basis1}$-$\eqref{sup_basis2}$. Imposing $\widetilde{h}_j$, $\widetilde{\gamma}_{kl}$ to be time-independent, we end up with a set of coupled first-order differential equations that determines the functions $f_j(t)$, $g_{kl}(t)$. 
Unfortunately, there is no obvious relation between  these new functions and the functions $\lambda_j(t),\, \pi_{kl}(t)$ that appear in Eq.~(\ref{param_dyn_map}).
Notice that since the choice of the auxiliary frame $\widetilde{\cal S}$ and of the initial conditions $W_{t_0}$ is arbitrary (and generally dictated by having a simplified set of differential equations), it may lead to an unphysical generator. However, if $\widetilde{\Li}$ is chosen to be a physical Liouvillian, then 
Eq.$\eqref{rot_frame}$ is the result of a generalised rotating frame transformation. We argue that the existence of a rotating frame transformation is not rare. In particular, one can always set $\widetilde{\Li}=0$ and $W_{t=t_0}=\Id_n \otimes \Id_n$ so that $\Lambda_{t,t_0}= W_t^{-1}$, as follows from Eq.$\eqref{rot_frame}$. This example has a clear physical interpretation: since the state of the system in $\widetilde{\cal S}$ is not evolving, the map $W_t$ generates a reversed time evolution. 

The rotating frame technique can also be viewed as an alternative to other techniques such as the coherence vector formalism (see e.g. \cite{Alicki-Lendi, Rau-2002,Rau-2003,Rau-2005}) where the time-dependent Lindblad equation is cast in the form of a Bloch equation for the components of the reduced density matrix. For completeness, we review the coherence vector formalism in  appendix \ref{coherence} and we apply it to the solution of a single qubit open quantum system, presented in the Sec. \ref{2lvl}.
 In the following paragraph, we shall consider the rotating frame technique applied to a periodic driving. In this case, the method leads to exact results that go beyond those obtained by means of other known techniques. 

\subsection{Periodic driving}\label{periodic_driving}

Let us discuss more in detail the case in which the time-dependent Lindblad equation is generated by a periodic Liouvillean $\Li_t=\Li_{t+\T}$ with a period $\T$. This problem has been analysed in many recent works, see e.g. \cite{per1,per2,per3,per4,per5,per6,per7,per8,per9,Prosen2018}, with the aim of extending the Floquet theory \cite{review_Floq} to the case of periodically driven open quantum systems. The dedicated literature mainly focus on the case of high-frequency driving where one may face the problem perturbatively \cite{Dai-16,Dai-17} with the use of the Magnus expansion, see e.g. \cite{review_magnus_exp}. Here, we show that within the rotating frame technique, one is able to derive an exact Floquet description of the Lindblad evolution $\eqref{Lind_eq}$, avoiding  problems of convergence \cite{Casas01,Moan2008} and of lack of CPTP properties \cite{Haddadfarshi-15} related to the Magnus expansion. Nonetheless, we mention that high-frequency perturbative results can be obtained within our framework and they generalise those of \cite{Rahav03,Goldman-PRX-14} to the open system's case. For more information, see  appendix \ref{high_freq}.


First, we require that the map $W_t$ introduced in Eq.$\eqref{W_map}$ also be periodic i.e., that the coordinates $f_j(t)=f_j(t+\T)$ and $g_{kl}(t)=g_{kl}(t+\T)$ are periodic functions. Then, by evaluating the dynamical map in Eq.$\eqref{rot_frame}$ at times $t=t_0+m\T$, for integer $m$, we obtain
\be\label{strob_map}
\left(\Lambda_{t,t_0}\right)\vert_{t=t_0+m\T} = W_{t_0}^{-1}\, e^{\widetilde{\Li} \, m\T} \, W_{t_0} = e^{\Li_F(t_0) \, m\T} 
\ee
where we introduced the Floquet generator
\be\label{Floq_Li}
\Li_F(t_0)= W_{t_0}^{-1} \, \widetilde{\Li} \, W_{t_0}\,.
\ee
The dynamical map in Eq.$\eqref{strob_map}$ describes a stroboscopic evolution whereas, introducing the micromotion operator
\be\label{kick}
\mathscr{K}_{t,t_0}= W_t^{-1} \, W_{t_0}\,,
\ee
one is able to analyse the evolution at any times $t\geq t_0$. We specify that the Floquet-Liouvillean $\Li_F$ in Eq.$\eqref{Floq_Li}$ is not unique and depends on the choice of the auxiliary frame. However, one can  set 
the  $\widetilde{\cal S}$-parameters $\widetilde{h}_j$, $\widetilde{\gamma}_{kl}$ in Eq.$\eqref{aux_Li2}$ to be equal to the time averages of the ${\cal S}$-parameters in Eq.$\eqref{Li}$: 
\be\label{averages}
\widetilde{h}_j=\overline{h}_j\,, \quad \widetilde{\gamma}_{kl}=\overline{\gamma}_{kl}
\ee
with $\overline{w}\equiv\T^{-1} \int_0^{\T}\,  dt\, w(t)$ for a generic function of time $w(t)$. Requiring $\eqref{averages}$, we fix the auxiliary frame $\widetilde{\cal S}$ and consequently $\Li_F$. Notice that the generator $\widetilde{\Li}$ in Eq.$\eqref{aux_Li}$ and $\Li_F$ are related through a similarity transformation $\eqref{Floq_Li}$ and so they share the same spectrum. This means that requiring $\eqref{averages}$ we are guaranteed that $\Li_F$ is a physical object.

As a consequence of the algebraic structure, the Floquet generator in Eq.$\eqref{Floq_Li}$ admits a decomposition:
\be
\Li_F(t_0)=\sum_{j=1}^{n^2-1} h^F_j(t_0) \, {\cal H}_j + \sum_{k,l=1}^{n^2-1} \gamma^F_{kl}(t_0)\, {\cal D}_{kl}\,,
\ee
in the superoperator basis of $\eqref{sup_basis1}$-$\eqref{sup_basis2}$. Here, the set of Floquet parameters $h^F_j$, $\gamma^F_{kl}$ are defined through the BCH expansion of the r.h.s. of Eq.$\eqref{Floq_Li}$.

Notice that it is not always possible to find a periodic solution to the set of non-linear first-order differential equations defining $W_t$ \cite{Li03}. This means that a periodic Liouvillean $\Li_t$ cannot be always expressed in terms of a Floquet theory. We shall not investigate the connection between the existence of periodic solutions and the markovianity of the Floquet evolution, leaving it to further developments. For our purposes, we assume the existence of a periodic solution and we provide a method to built a Floquet-Liouvillean $\Li_F$. We address the reader to  Ref.~\cite{Schnell18} where the existence of $\Li_F$ is discussed and some markovianity tests are proposed (see also \cite{Wolf08,Cubitt12}). 
\section{Single qubit example}\label{2lvl}

For concreteness, we shall focus now on the case $n=2$.
We assume that the system is described by a time-dependent Hamiltonian of the form 
\be\label{2lvl_Ham}
H_t= -\frac{1}{2}\Omega(t) \, \sigma_3
\ee
where $\sigma_3$ is the diagonal Pauli operator. 
In addition, we add three elementary dissipative processes described by the jump operators 
\be\label{2lvl_jumps}
L_1= \sigma_+\,, \quad L_2=\sigma_-\,, \quad L_3= \sigma_3
\ee
where $\sigma_{\pm}= (\sigma_1\pm i \sigma_2)/2$ are ladder combinations of Pauli operators. 
The first two represent a finite temperature amplitude damping, whereas the last one represents a dephasing in the $\sigma_3$ basis. 
The full master equation is therefore taken as 
\be\label{Lindblad_2lvl}
\frac{d}{dt} \rho_t=\frac{i}{2}\Omega(t)[\sigma_3,\rho_t] + \sum_{j=\pm,3} \Gamma_j(t) (\sigma_j \rho_t\sigma_j^{\dagger}-\frac{1}{2}\{\sigma_j^{\dagger}\sigma_j,\rho_t\})
\ee

The natural operator basis $F_i$ in this case is given by the normalised Pauli matrices $\sigma_i/\sqrt{2}$ ($i=0,1,2,3$) where $\sigma_0$ is the identity $\Id_2$. 
Then, writing the master equation in vectorised form, as in  Eqs.~(\ref{Lind_vec}) and (\ref{Li}), we find 
\be\begin{split}\label{Liouv_2lvl_2}
\Li_t= &-\frac{\Omega(t)}{\sqrt{2}}\, {\cal H}_3 +\alpha(t)(\D_{11}+\D_{22}) \\
&+ i\beta(t) (\D_{12}-\D_{21}) + 2\Gamma_3(t) \D_{33}
\end{split}\ee
where ${\cal H}_i$ and $\D_{ij}$ are defined in Eqs.~(\ref{sup_basis1}) and (\ref{sup_basis2}).
Moreover, we have defined 
$\alpha(t)=(\Gamma_+(t)+\Gamma_-(t))/2$ and $\beta(t)=(\Gamma_+(t)-\Gamma_-(t))/2$. 
At this point, it is useful to introduce the combinations of superoperators
\be
{\cal D}_{\uparrow\,; \, \downarrow}=\frac{1}{2}(\D_{11}+\D_{22}\mp i \D_{12} \pm \D_{21})
\ee
in terms of which the Liouvillean in Eq.$\eqref{Liouv_2lvl_2}$ has a diagonal structure. 
Moreover, the set of superoperators $\{ {\cal H}_3, \D\p, \D\m, \D_{33}\}$ form a simple closed subalgebra having only
\be
[\D\p, \D\m]= \D\p-\D\m
\ee
as a non-zero commutator. 

We shall now provide an exact solution for the time-dependent Lindblad equation $\eqref{Lindblad_2lvl}$ using the rotating frame technique of Sec.\ref{rotating_frame}. We design a map $W_t$ to $\widetilde{\cal S}$ in the form
\be
W_t= e^{f(t) {\cal H}_3} \, e^{g_1(t) \D\p} \, e^{g_2(t) \D\m} \, e^{g_{3}(t) \D_{33}}
\ee
and from Eqs.$\eqref{aux_Li}$, $\eqref{aux_Li2}$ we obtain the set of coupled first-order differential equations:
\begin{subequations}\label{eq_aux}
\be\label{eq_aux_unitary}
\frac{d}{dt}f(t)- \frac{1}{\sqrt{2}}  (\Omega(t) - \widetilde{\Omega})= 0\, ;
\ee
\be\label{eq_aux_diss3}
\frac{d}{dt} g_3(t) + 2(\Gamma_3(t)- \widetilde{\Gamma}_3 )=0\,;
\ee
\begin{widetext}
\be\label{eq_aux_diss1}
\widetilde{\Gamma}_+= \frac{d}{dt} g_1(t) + \left( 1- e^{-g_1(t)}\right)\, \frac{d}{dt} g_2(t) +\left(1+e^{-(g_1(t)+g_2(t))}-e^{-g_1(t)}\right) \Gamma_+(t)+ \left(1-e^{-g_1(t)}\right) \Gamma_-(t)\,;
\ee
\be\label{eq_aux_diss2}
\widetilde{\Gamma}_-=e^{-g_1(t)} \left( \frac{d}{dt} g_2(t) + \Gamma_+(t) + \Gamma_-(t)\right) - e^{-(g_1(t)+g_2(t))} \Gamma_+(t) \,,
\ee
\end{widetext}
\end{subequations}
where the values at $t=t_0$ of the functions $f$, $g_1$, $g_2$ and $g_3$ are specified through the choice of the initial state $\ket{\widetilde{\rho}_{t_0}}$ in $\widetilde{\cal S}$.   Eqs. $\eqref{eq_aux_unitary}$ and $\eqref{eq_aux_diss3}$ are readily solvable while, taking the sum and the difference of $\eqref{eq_aux_diss1}$-$\eqref{eq_aux_diss2}$, we obtain
\be\begin{split}
r(t)&\equiv g_1(t)+g_2(t)\\
&=- \int^t dt^{\prime} \, \left((\Gamma_+(t^{\prime}) -\widetilde{\Gamma}_+)+(\Gamma_-(t^{\prime}) -\widetilde{\Gamma}_-)\right)
\end{split}\ee
and
\be\label{Y_eq}
\frac{d}{dt}y(t) + (\Gamma_+(t)+\Gamma_-(t)) y(t) - \Gamma_+(t) -e^{r(t)} \widetilde{\Gamma}_-=0\,,
\ee
where $y(t)\equiv \exp(g_2(t))$. 

For a periodic driving, we can conveniently require the condition $\eqref{averages}$ and look for a periodic solution of  Eqs.$\eqref{eq_aux}$. In this case, one finds the Floquet parameters $\Omega^F=\overline{\Omega}$, $\Gamma_3^F=\overline{\Gamma}_3$ and
\be\label{Gamma_Floq}
\Gamma_{\pm}^F(t_0) = \overline{\Gamma}_{\pm} \mp \delta\Gamma(t_0)
\ee
with
\be\label{Gamma_Floq2}
\delta\Gamma(t_0) =(1-y(t_0)) \overline{\Gamma}_+ + ( e^{r(t_0)}-y(t_0)) \overline{\Gamma}_-\,.
\ee

\subsection{Counter-oscillating polarisers}\label{counter_oscillating_case}

\begin{figure}
 \includegraphics[scale=0.4]{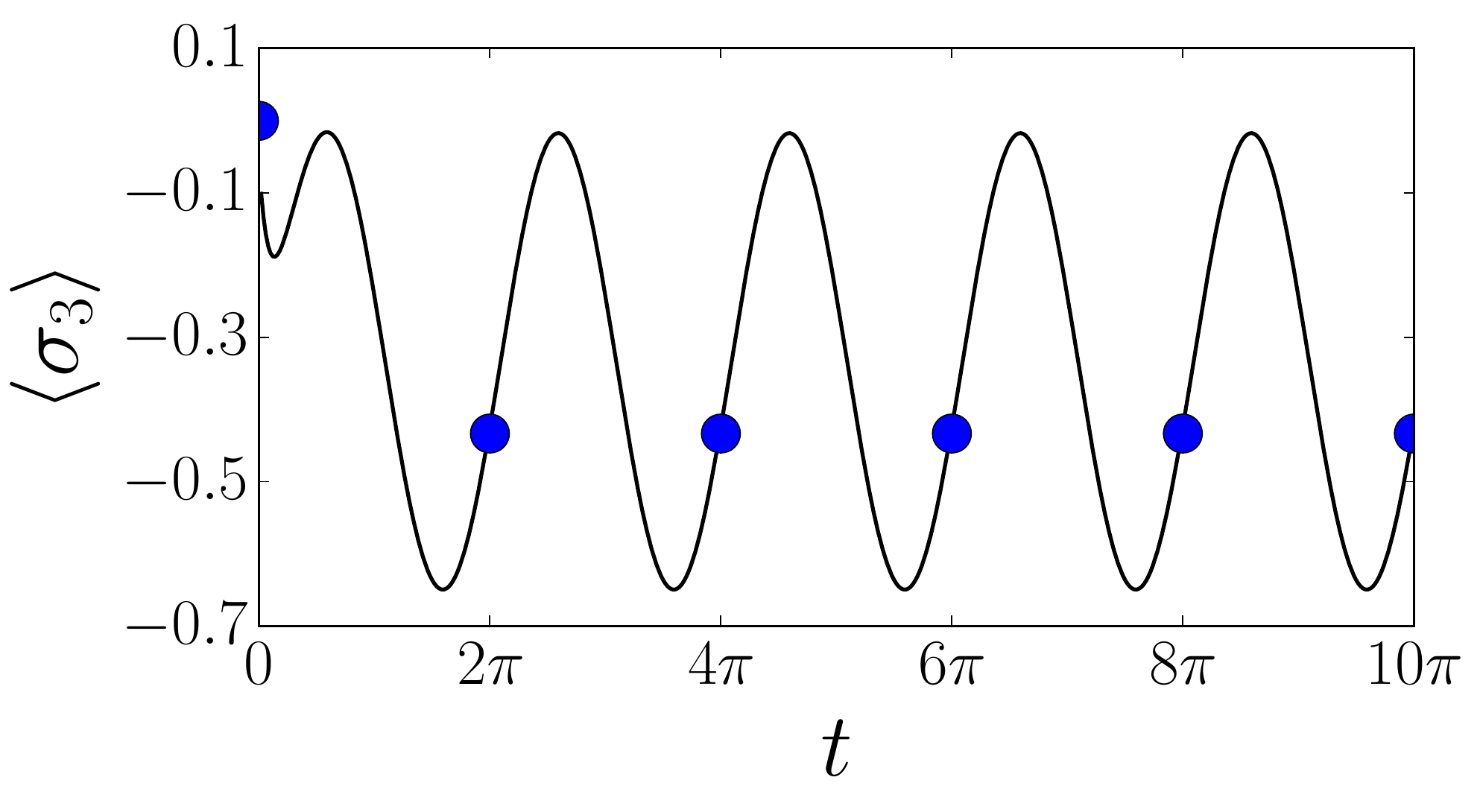}
 \caption{Time evolution of the spin magnetisation $\braket{\sigma_3}=\tr(\rho_t\,\sigma_3)$ for the two-levels open quantum system in Eq.$\eqref{Lindblad_2lvl}$ with counter-oscillating polarisers $\Gamma_+=2.0+ 0.5 \sin(t)$, $\Gamma_-=3.0- 0.5 \sin(t)$ and $\Gamma_3=0$, $\Omega=\sqrt{2}(1-\cos(t))$. The numerical exact time evolution ({\it full line}) is compared with the stroboscopic evolution ({\it dots}) given by Eq.$\eqref{counter_shift}$. The system is prepared at $t=0$ in a state $\rho_0=\Id_2/2$.}\label{spin1}
 \end{figure}

As a first illustration, we consider the case of a periodic driving with polarisers
\be\label{counter_pol}
\Gamma_{\pm}(t)= \overline{\Gamma}_{\pm} \pm A\sin(\omega t)
\ee
where $|A|\leq \overline{\Gamma}_{\pm}$ and $\omega\equiv 2\pi/\T$. For this setting, the solution of Eq.$\eqref{Y_eq}$ is given by
\be
y(t)=1+\frac{A}{\Gamma^2 + \omega^2} (\Gamma\sin(\omega t) -\omega \cos(\omega t))\,,
 \ee
 $\Gamma=\overline{\Gamma}_++\overline{\Gamma}_-$, and leads to a Floquet shift of the polarisers
 \be\label{counter_shift}
 \delta\Gamma(t_0)=A\omega\Gamma/(\Gamma^2+\omega^2)\,.
 \ee
 In Fig.\ref{spin1} we show the time evolution of the magnetisation of a single qubit subject to counter-oscillating polarisers $\eqref{counter_pol}$ for a specific choice of the parameters in Eq.$\eqref{Lindblad_2lvl}$.
 
 \subsection{Incoherent driving}\label{incoherent_driving}
\begin{figure}
 \includegraphics[scale=0.55]{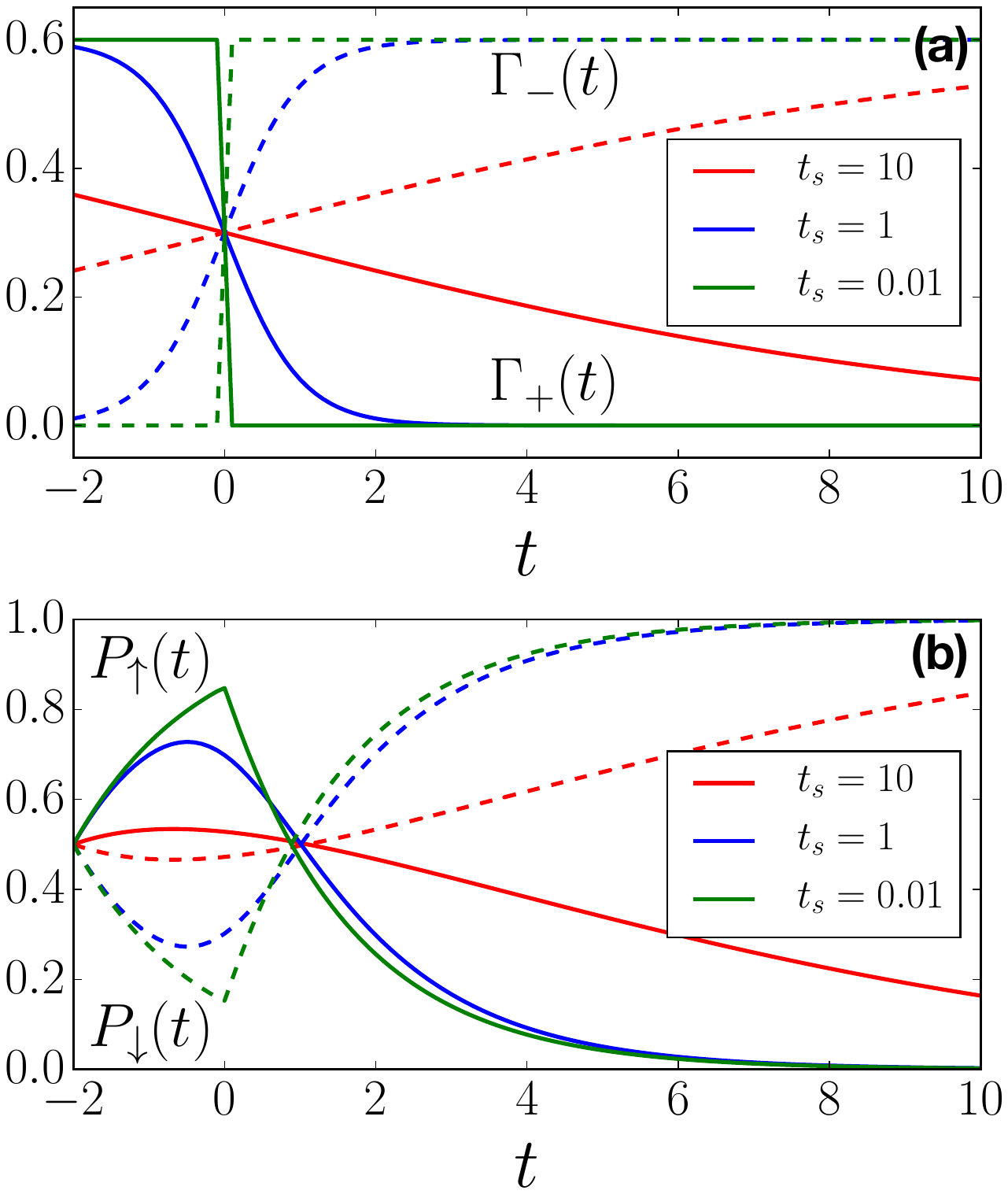}
 \caption{{\bf(a)} The incoherent driving is made through the polarisers in Eq.$\eqref{polarisers_incoherent}$. In the plot we show the time evolution of $\Gamma_+$ ({\it full line}) and $\Gamma_-$ ({\it dashed line}) for three values of the time scale $t_s$, ranging from an adiabatic to a quench regime. The amplitude is set to the value $A=0.3$. {\bf(b)} Time evolution of the population $\eqref{pop}$ of the excited (ground) state in {\it full line} ({\it dashed line}) of a single qubit subject to the polarisers in Eq.$\eqref{polarisers_incoherent}$. We prepare our system at $t_0=-2$ in a high-temperature state $\rho_{t_0}=\Id_2/2$ with $A=0.3$ and $\Omega=1.0$. We clearly see that the system changes its target state when the amplitudes of the two lasers cross each other. This effect becomes sharper for short time scale $t_s$. In particular, in the quench regime ({\it green line}) the population present a cusp point at the crossover related to the sudden switch of the laser direction.}\label{incoherent}
 \end{figure}

Next we provide an example of a non-periodic driving for the Lindblad evolution in Eq.$\eqref{Lindblad_2lvl}$ of a single qubit. In particular, we shall consider the case where $\Gamma_3=0$, $\Omega$ 
 is kept constant and where the system is incoherently driven by the polarisers
\be\label{polarisers_incoherent}
\Gamma_+(t)=A ( 1 - \tanh(t/t_s))\,, \quad \Gamma_-(t)=A ( 1 + \tanh(t/t_s))
\ee
that we plot in Fig.\ref{incoherent}a for different values of the time scale $t_s$ of the driving. 

In other words, the qubit  feels the resulting effect of the two competing drives and thus is always trying to follow a target state that is changing in time. 
From the Eqs.$\eqref{param_dyn_map}$,$\eqref{non_unitary_int}$, it follows that the time evolution is generated by the dynamical map
\be
\Lambda_{t,t_0}= \exp(\frac{-\Omega}{\sqrt{2}} (t-t_0) \, {\cal H}_3)\, \exp(\pi_1(t) \, \D\p) \, \exp(\pi_2(t) \, \D\m)
\ee
where the functions $\pi_1$, $\pi_2$ are defined through 
\begin{subequations}\label{eq_incoherent}
\be
\frac{d}{dt} \pi_1(t)+ (1-e^{-\pi_1(t)})\frac{d}{dt} \pi_2(t)=\Gamma_+(t)\,,
\ee
\be
 e^{-\pi_1(t)} \, \frac{d}{dt} \pi_2(t) = \Gamma_-(t)
\ee
\end{subequations}
with initial conditions $\pi_{1,2}(t_0)=0$ and formal solutions
\begin{subequations}
\be
\pi_2(t)=\log\left(1+\int_{t_0}^t dt^{\prime} \, e^{2A\; (t^{\prime}-t_0)} \, \Gamma_-(t^{\prime})\right)\,,
\ee
\be
\pi_1(t)=-\pi_2(t)+2A\; (t-t_0).
\ee
\end{subequations}
Defining the projectors $P\p=\ket{\uparrow}\bra{\uparrow}$ on the excited state and $P\m=\ket{\downarrow}\bra{\downarrow}$ on the ground state, one can then follow the time evolution of the populations 
\be\label{pop}
P\p(t)=\tr\left(\ket{\uparrow}\bra{\uparrow}\, \rho_t\right)=  \tr\left( \ket{\uparrow}\bra{\uparrow}\,\Lambda_{t,t_0}\, \rho_{t_0} \right)
\ee
 and similar for $P\m(t)$, as shown in Fig.\ref{incoherent}b.
 

%

\section{Quantum heat-engines}\label{quantum_heat_engines} 

An interesting application of the Floquet-Lindblad formalism outlined in Sec.\ref{periodic_driving} regards the investigation of finite-time quantum heat-engines \cite{Alicki-79,Kosloff-Levy,Alicki2015,Rezek2006,Abah2014,Klaers2017a,Ronagel2014,Correa2014,Jaramillo2016,Samuelsson2017,DelCampo2014,Abah2016,Abah2017,Camati2016,Elouard2017,Cottet2017,Masuyama2017,Manzano2017b,Micadei2017,Manzano2017,Scopa-18}. The operation of these devices rely on the existence of a stroboscopic steady state $\rho^{ss}(t)$, which is time-periodic and towards which the system converges during the early stages of the time evolution. Once the dynamics becomes periodic, the engine's operation can be characterized through the time evolution of thermodynamic quantities such as the internal energy $E$ or the heat current $\delta{\cal Q}$. 
In  appendix \ref{quantum_therm} we review some definitions and terminology of basic quantum thermodynamics, needed in the following. For a wider and more complete understanding we address the reader to  Refs.\cite{Anders17,Kosloff_rev,AlickiKossloff18}.

Although it is easy to identify the basic elements of an abstract quantum heat-engine, understanding what is the state of the system within the cyclic dynamics and if this state can be eventually reached
are non-trivial tasks. 
In Ref.~\cite{Scopa-18} we have shown that a  clear answer to these questions can be provided by the Floquet generator $\Li_F$ in Eq.$\eqref{Floq_Li}$: the convergence to a cyclic evolution is directly associated to the eigenvalues of $\Li_F$ while the state of the system within the cycle is simply the instantaneous eigenstate of $\Li_F(t)$, with $t$ as a parameter [cf. Eq.~(\ref{Floq_Li})].
That is, 
\be\label{steady_state}
\Li_F(t) \ket{\rho^{ss}_{t}}=0\,.
\ee
Thus, all relevant properties of the heat engine's finite-time operation are contained in $\Li_F(t)$.

\subsection{Two-levels quantum heat-engine}\label{2lvl_engine}

\begin{figure}
\includegraphics[scale=0.3]{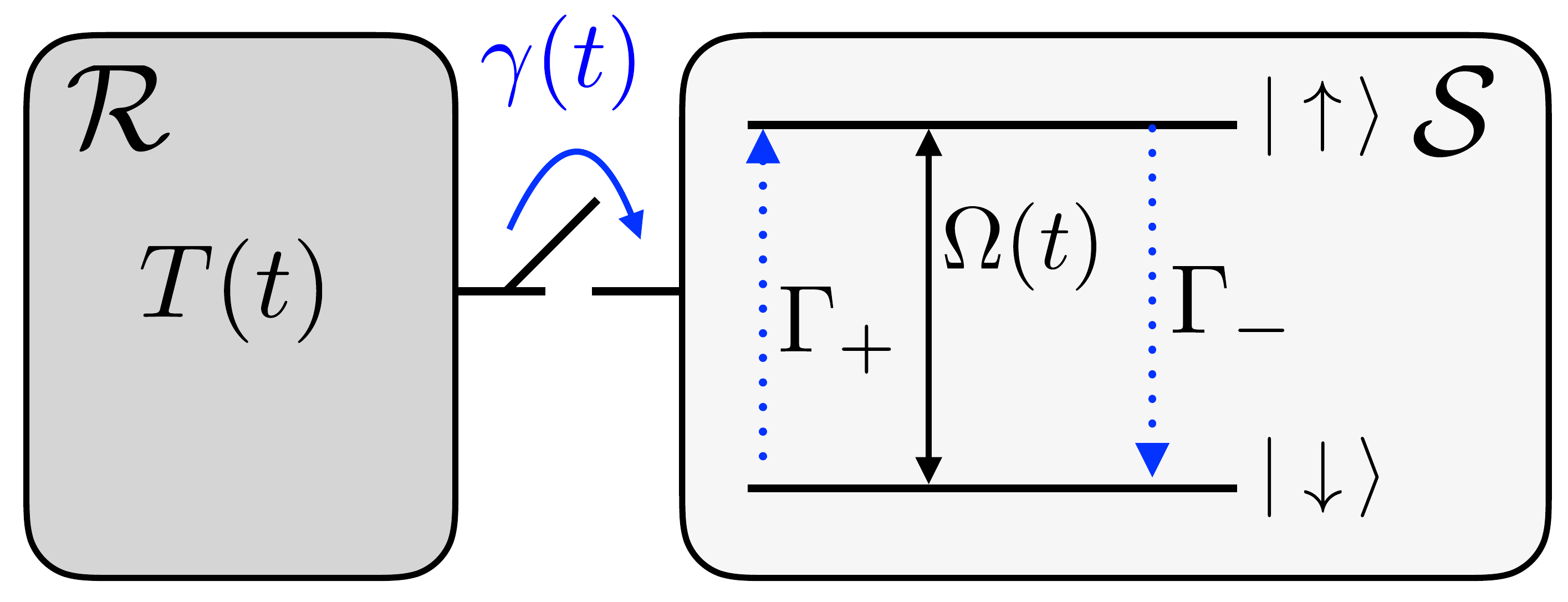}
\caption{Illustration of a two-levels quantum heat engine. The Hamiltonian parameter $\Omega(t)$ tunes the energy gap between the ground state $\ket{\uparrow}$ and the excited level $\ket{\downarrow}$. The thermal bath is modelled as in Eq.$\eqref{detailed_balance}$ with a couple of polarisers $\Gamma_+$, $\Gamma_-$. The interactions of the system ${\cal S}$ with the thermal bath ${\cal R}$ are controlled by the interruptor $\gamma$.}\label{2lvl_engine}
\end{figure}

We now apply the general results presented in Sec.\ref{quantum_heat_engines} for the concrete design of a quantum heat-engine operating with a working fluid composed of a single qubit \cite{Geva-92}. In particular, we consider the Lindblad evolution in Eq.$\eqref{Lindblad_2lvl}$ with $\Gamma_3=0$ and with polarisers satisfying a detailed balance condition
\be\label{detailed_balance}
\frac{\Gamma_-}{\Gamma_+}= \exp(\Omega(t)/T(t))\,, \quad \forall t\geq t_0
\ee
so that their combined effect mimic the relaxation properties induced by a thermal bath ${\cal R}$ with time-dependent temperature $T(t)$. A convenient parametrisation of the thermal bath is
\be\label{thermal_bath_gamma}
\Gamma_+(t)=\gamma(t)\,n(t)\,, \quad \Gamma_-(t)=\gamma(t)(1-n(t))
\ee
with 
\be
n(t)=(1+ e^{\Omega(t)/T(t)})^{-1}
\ee
 and a control parameter $\gamma(t)$ that allows us to couple/uncouple the system with the bath at different strokes. An illustration of the $2$-levels quantum heat engine is given in Fig.\ref{2lvl_engine}.
 
Inserting Eqs.$\eqref{Gamma_Floq}$ and $\eqref{Gamma_Floq2}$ in Eq.$\eqref{steady_state}$, we obtain 
\be
\rho^{ss}_t=F_0 - \frac{\sqrt{2}}{\Gamma} \, \delta\Gamma(t) F_3
\ee
with $\Gamma=\overline{\Gamma}_++ \overline{\Gamma}_-$. The convergence towards $\rho^{ss}_t$ is dictated by the real part of the eigenvalues of $\Li_F$
\be\label{transient}
\text{spec}(\Li_F)=\left( 0,\, -\Gamma,\, -\frac{1}{2}(\,\Gamma\pm i2\sqrt{2}\; \overline{\Omega})\right)
\ee
from which we specificy the time scale $\tau$ of the transient regime to be $\tau= 1 /\Gamma$. The cyclic evolution of the internal energy $E$ then reads
\be\label{energy_F}
E(t) =\braket{H_t}=\Omega(t)  \frac{\delta\Gamma(t)}{\Gamma}\,,
\ee
$\braket{\bullet}\equiv \tr(\rho^{ss}_t \, \bullet )$, and similarly we obtain the results
\be
{\cal P}(t)=\left\langle\frac{\de}{\de t} H_t\right\rangle=\left(\frac{d}{dt} \,\Omega(t) \right)\, \frac{\delta\Gamma(t)}{\Gamma}\,, 
\ee\be
 \delta{\cal Q}(t)=\braket{D^{\star}(H_t)}=2\Omega(t)\, \frac{\Gamma_+^F(t)\,\Gamma_-^F(t)}{\Gamma}
 \ee
for the power output ${\cal P}$ and the heat flow $\delta{\cal Q}$ respectively.

\subsubsection{Carnot cycle}\label{carnot_2lvl}
\begin{figure}
\includegraphics[scale=0.6]{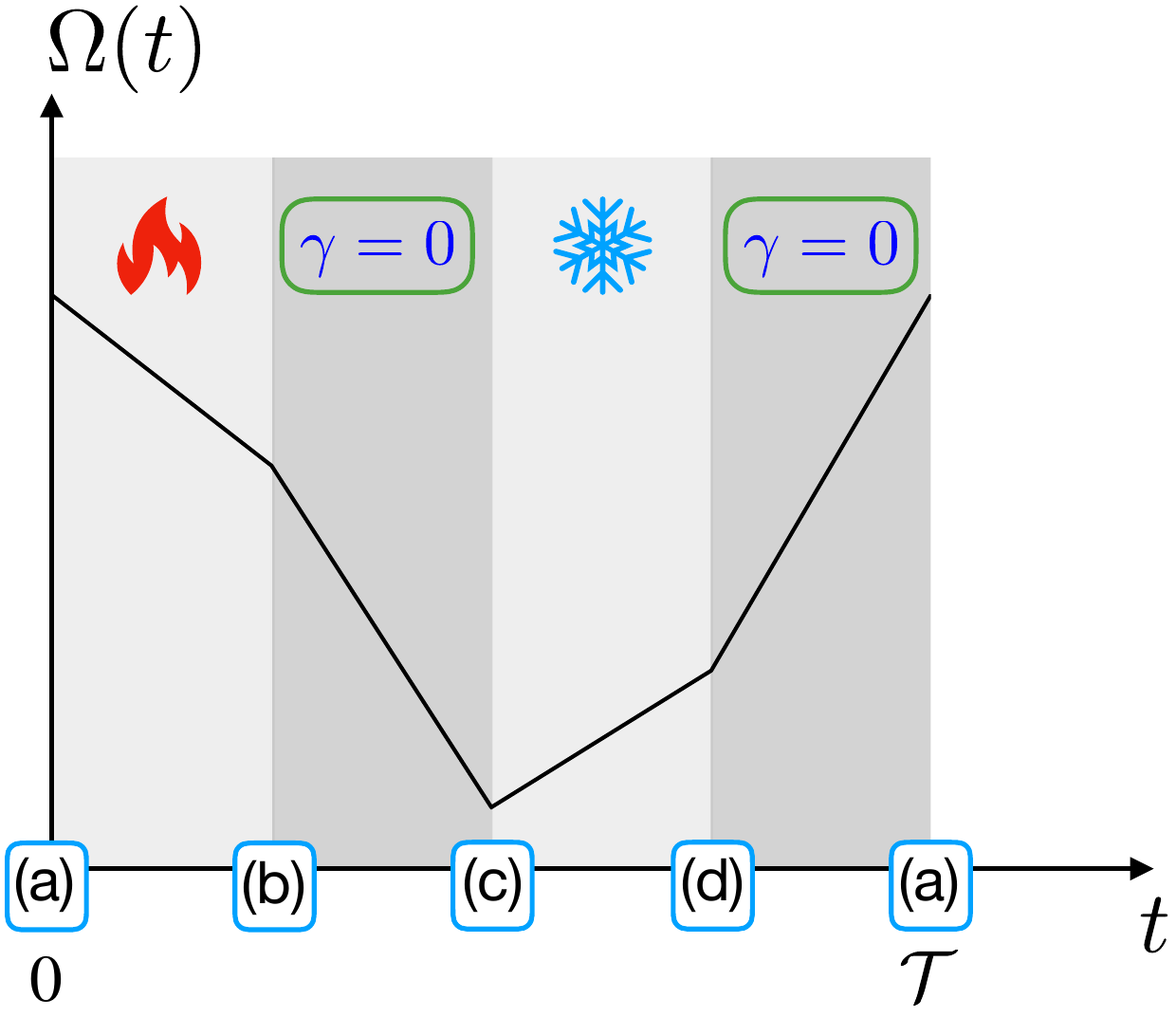}\\
\caption{A sketch of a Carnot cycle where the working parameter $\Omega$ is varied in time as a linear piecewise function.}\label{carnot_sketch}
\end{figure}
To illustrate a concrete application of the previous results, we shall consider the operation of a finite-time $2$-level Carnot engine. The cycle is made of four strokes, which we assume have an equal duration $\T/4$:
\begin{enumerate}
\item {\bf(ab)} Hot isothermal expansion at $T\hot$
\item {\bf(bc)} Isentropic expansion
\item {\bf(cd)} Cold isothermal compression at $T\cold$
\item {\bf(da)} Isentropic compression
\end{enumerate}
For our specific model, expansions (compressions) mean decreasing (increasing) the level spacing through the parameter $\Omega$. We shall consider a protocol where $\Omega(t)$ varies in a linear piecewise fashion, as sketched in Fig.\ref{carnot_sketch}. The control function $\gamma(t)$ in Eq.$\eqref{thermal_bath_gamma}$ is set to
\be
\gamma(t)= \left\{ \,1,\, 0 ,\, 1 ,\, 0\, \right\}\,,
\ee
since the system evolves unitarily in the second and in the fourth stroke. Finally, the time-dependence of $T(t)$  is given by
\be\label{temperature}
T(t)=\left\{ T\hot,\, - \,, T\cold,\, -\,\right\}\,, \quad T\hot>T\cold
\ee
where the notation "$-$" means that the specific value of the temperature in that interval is not physically relevant since the thermal bath is detached from the system.\\

\subsubsection{Quasi-static reversibility limit}
For large enough values of the driving period $\T \gg1$, the evolution of the quantum heat-engine can be investigated by using standard equilibrium statistical mechanics. 
During the isothermal strokes the expectation value of the energy reads
\be\label{thermal_qs}
E(t)=\braket{H_t}_{eq}= \frac{\Omega(t)}{2} \, \tanh(\frac{\Omega(t)}{2T})
\ee 
where $\braket{\bullet}_{eq}\equiv \tr(\rho^{eq} \, \bullet)$ is the average computed with the thermal equilibrium density matrix $\rho^{eq}$ at a fixed temperature $T$:
\be
\rho^{eq}= \frac{1}{2\cosh(\Omega(t)/T)} \begin{pmatrix} e^{-\Omega(t)/2T} & 0 \\ 0 & e^{\Omega(t)/2T}\end{pmatrix}\,.
\ee
On the other hand, in the isentropic strokes the evolution is unitary and any variation in the internal energy is due to a variation of the energy separation of the two levels
\be\label{isentropic_qs}
E(t_2)= \frac{\Omega(t_2)}{\Omega(t_1)}\, E(t_1)\,,
\ee
as follows from the adiabatic theorem in standard quantum mechanics. We can see that depending on the driving protocol being used, the cycle may not have a reversible quasi-static limit. The reason is that, if by the end of the isentropic strokes ({\bf c} and {\bf a}) the value of the energies in Eq.$\eqref{isentropic_qs}$ are not the same to those at thermal equilibrium in Eq.$\eqref{thermal_qs}$ with a hot and a cold temperatures, then a dissipation inevitably will take place.
The condition for the existence of a quasi-static reversibility condition is therefore obtained by imposing that \cite{Scopa-18,Sekimoto2000,Leksha2018} 
\be\label{quasi_static_rev}
\frac{T\cold}{T\hot}=\frac{\Omega({\bf c})}{\Omega({\bf b})}=\frac{\Omega({\bf d})}{\Omega({\bf a})}\,.
\ee
These conditions are usually referred to as {\it quasi-static reversibility}.

\subsubsection{Finite-time operation}

We now turn to the finite time operation. With Eq.$\eqref{quasi_static_rev}$ in mind, we choose the unitary parameter to vary linearly in the four strokes, as:
\be\label{Omega_carnot}
\Omega(t)=\begin{cases} \frac{4(\Omega_b-\Omega_a)}{\T} t + \Omega_a \\  \frac{4(\Omega_c-\Omega_b)}{\T} t + 2\Omega_b-\Omega_c\\
 \frac{4(\Omega_c-\Omega_d)}{\T} t + 3\Omega_c -2\Omega_d \\  \frac{4(\Omega_a-\Omega_d)}{\T} t + 3\Omega_a -4\Omega_d \end{cases}
 \ee
with $\Omega_c=(T\cold/T\hot)\Omega_b$ and $\Omega_d=(T\cold/T\hot)\Omega_a$. \\

\begin{figure}
\includegraphics[scale=0.55]{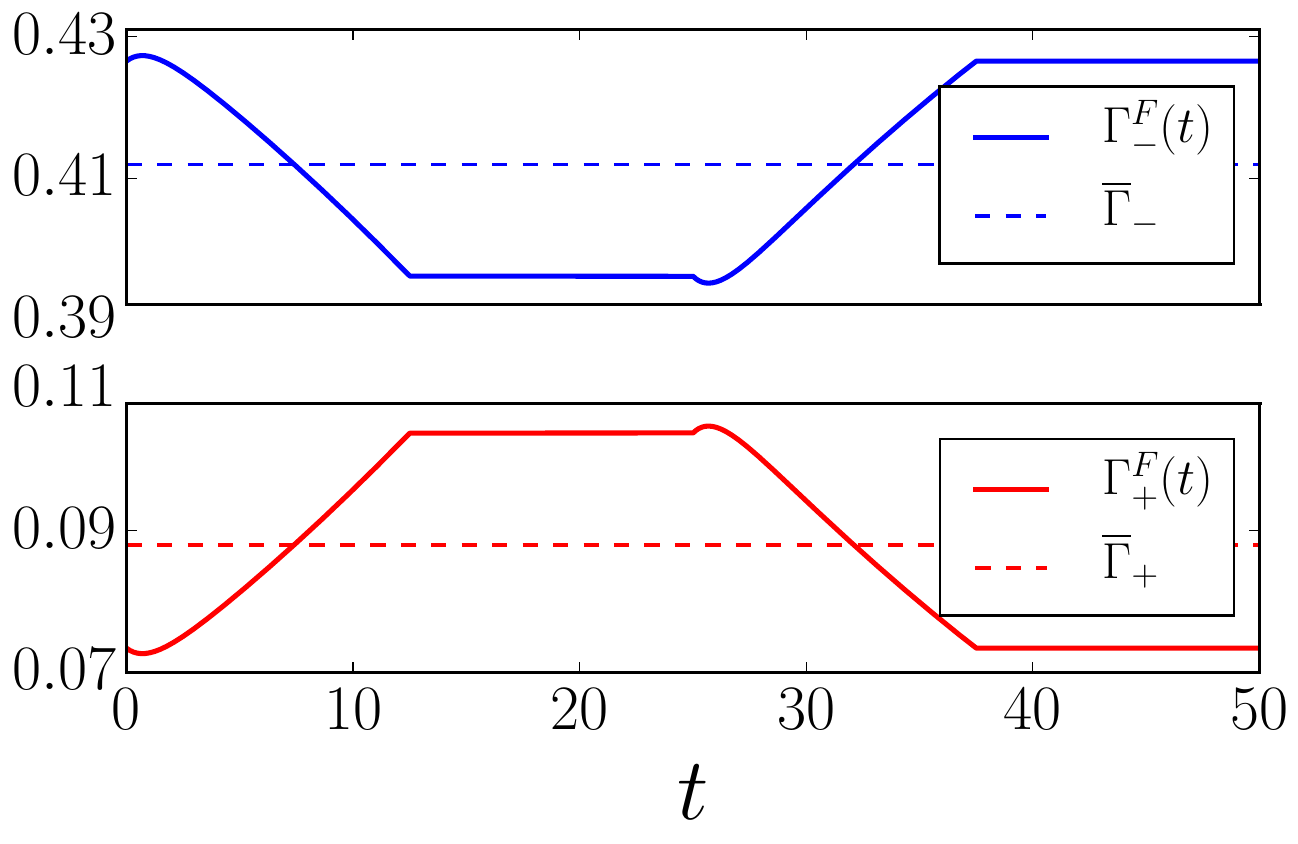}
\caption{Carnot engine. Numerical results for the Floquet parameters in Eqs.$\eqref{Gamma_Floq}$-$\eqref{Gamma_Floq2}$ obtained by solving Eq.$\eqref{Y_eq}$ with polarisers in Eq.$\eqref{thermal_bath_gamma}$ and energy level separation $\Omega$ in Eq.$\eqref{Omega_carnot}$, setting $\Omega_a=1.8$, $\Omega_b=1.3$ and baths' temperatures $T\hot=1.0$, $T\cold=0.5$. The driving period is $\T=50$. }\label{floq_gamma}
\end{figure}

\begin{figure}
\includegraphics[scale=0.4]{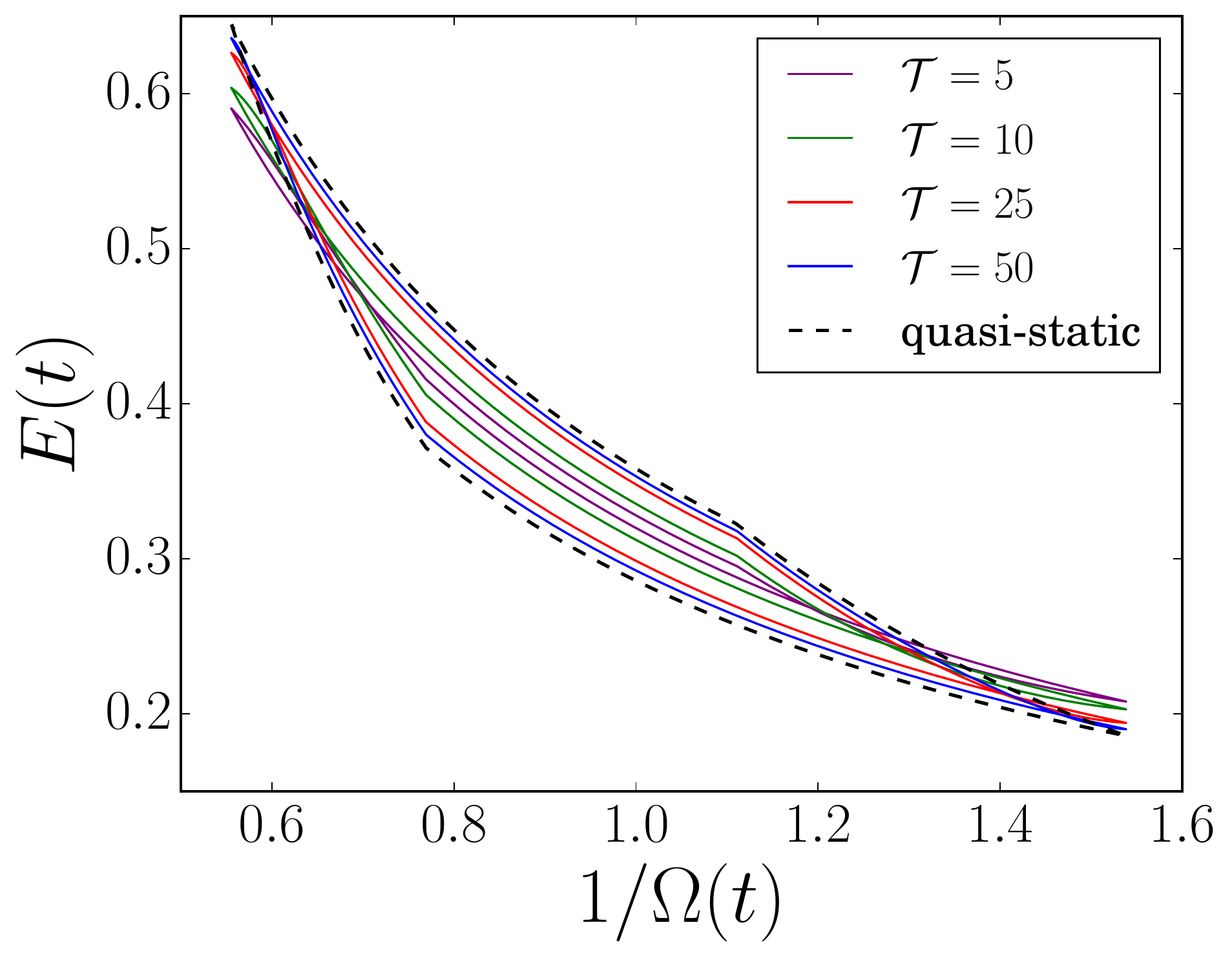}
\caption{Operation of a finite-time Carnot engine. The plot shows the internal energy $E(t)$ as function of the compression $1/\Omega(t)$. In the figure we select $\Omega$ as in Eq.$\eqref{Omega_carnot}$, setting $\Omega_a=1.8$, $\Omega_b=1.3$ and the thermal bath in Eq.$\eqref{thermal_bath_gamma}$ with $T\hot=1.0$, $T\cold=0.5$. We see that at large enough values of the driving period $\T$, the cycle matches its quasi-static limit.}\label{carnot_plot}
\end{figure}

\begin{figure}
\includegraphics[scale=0.4]{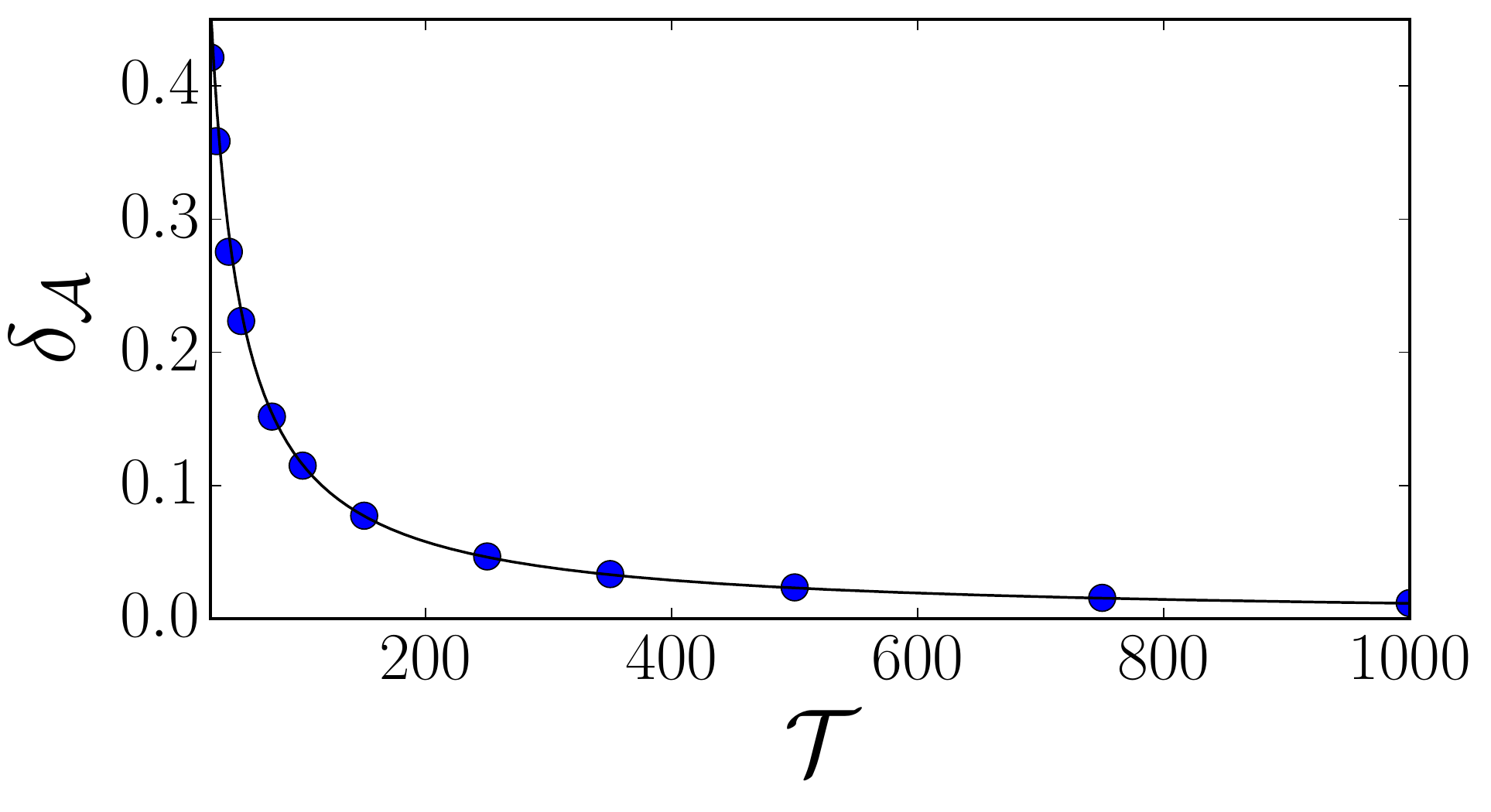}
\caption{A numerical estimation of the deviations of the cycle areas $\delta_{\cal A}$ in Eq.$\eqref{area_C}$ for the 2-levels Carnot engine in Fig.\ref{carnot_plot}. We see that the approach to the quasi-static limit is $\propto 1/\T$ with a proportionality constant $c=11.6$, extracted from our fitting datas.}\label{area}
\end{figure}

A numerical solution of  Eq.$\eqref{Y_eq}$ with $\Gamma_{\pm}$ set as in Eq.$\eqref{thermal_bath_gamma}$ allow us to determine the Floquet parameters $\Gamma_{\pm}^F$ in Eqs.$\eqref{Gamma_Floq}$-$\eqref{Gamma_Floq2}$. We show the results in Fig.\ref{floq_gamma}. From Eq.$\eqref{energy_F}$, one is then able to compute the cyclic evolution of the energy  $E(t)$ in the Carnot cycle that we plot in Fig.\ref{carnot_plot} for different values of the driving period $\T$. The limit $\T \gg 1$ is an important test of our results that has been well-recovered. All the deviations at finite-time operation from the quasi-static limit can be addressed to a non-equilibrium behavior of the device. The latter can be quantified through the deviations of the cycle area ${\cal A}$ from its quasi-static limit value ${\cal A}_{QS}$ :
\be\label{area_C}
\delta_{\cal A}= 1- \frac{\cal A}{{\cal A}_{QS}}\,.
\ee
A numerical estimation of $\delta_{\cal A}$ shows that the approach to the equilibrium quasi-static operation is $\delta_{\cal A}\sim {\cal O}(1/\T)$, see Fig.\ref{area}.\\

\begin{figure}
\includegraphics[scale=0.4]{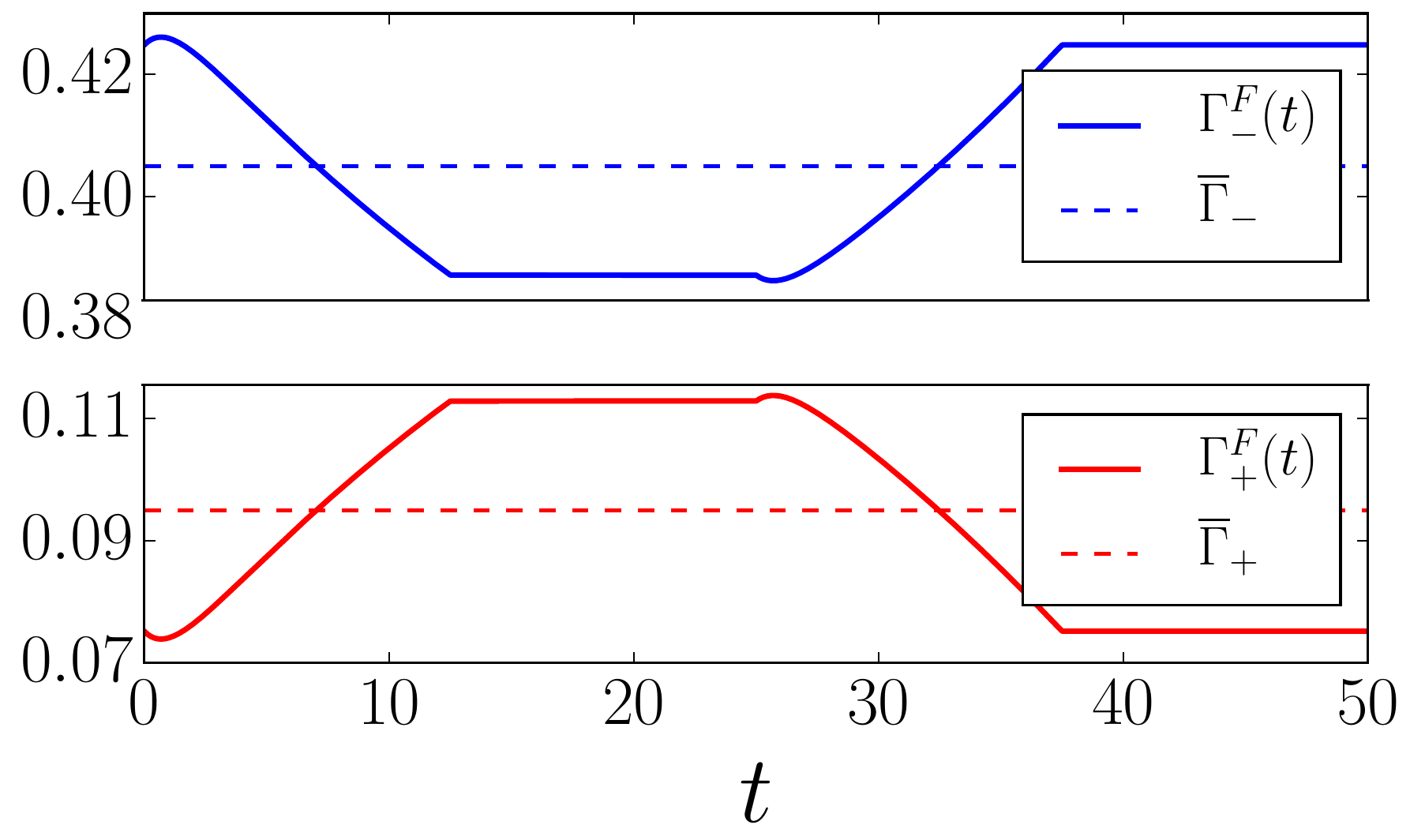}
\caption{Otto engine. Numerical results for the Floquet parameters in Eqs.$\eqref{Gamma_Floq}$-$\eqref{Gamma_Floq2}$ obtained by solving Eq.$\eqref{Y_eq}$ with polarisers in Eq.$\eqref{thermal_bath_gamma}$
and energy level separation $\Omega$ in Eq.$\eqref{Omega_otto}$, setting $\Omega_1=1.8$, $\Omega_2=1.3$. The temperature is varied as in Eq.$\eqref{temp_otto}$, requiring quasi-static reversibility conditions $\eqref{otto_QS}$ and chosing $T_a=1.0$, $T_b=1.5$. The driving period is $\T=50$. }\label{floq_gamma_otto}
\end{figure}

\begin{figure}
\includegraphics[scale=0.4]{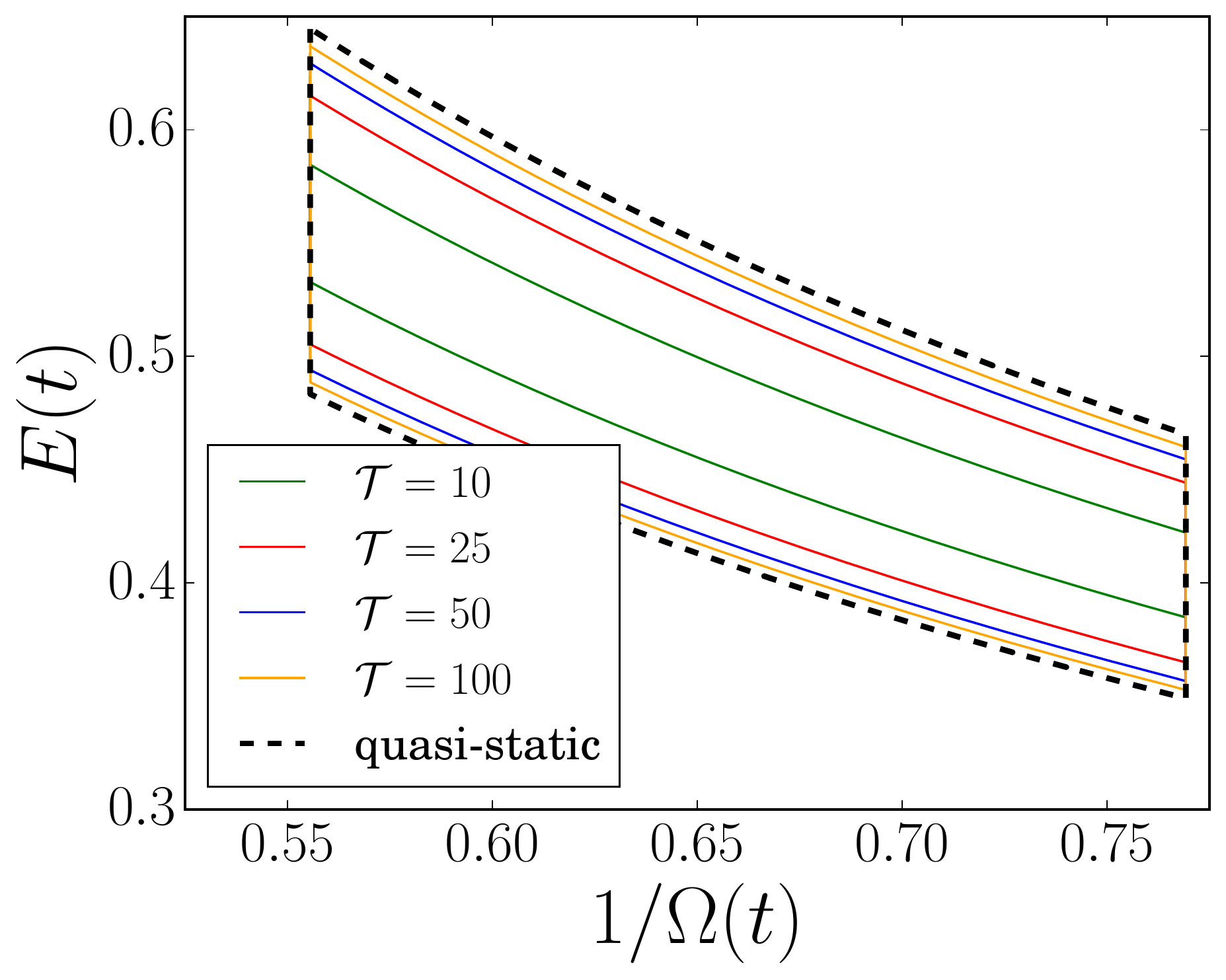}
\caption{Operation of a finite-time Otto engine. The plot shows the internal energy $E(t)$ as function of the compression $1/\Omega(t)$. In the figure we select $\Omega$ as in Eq.$\eqref{Omega_otto}$, setting $\Omega_1=1.8$, $\Omega_2=1.3$ and the thermal bath in Eq.$\eqref{thermal_bath_gamma}$ with temperature in Eq.$\eqref{temp_otto}$. We require quasi-static reversibility conditions $\eqref{otto_QS}$ and we chose $T_a=1.0$, $T_b=1.5$. We see that at large enough values of the driving period $\T$, the cycle matches its quasi-static limit.}\label{otto_plot}
\end{figure}

\begin{figure}
\includegraphics[scale=0.4]{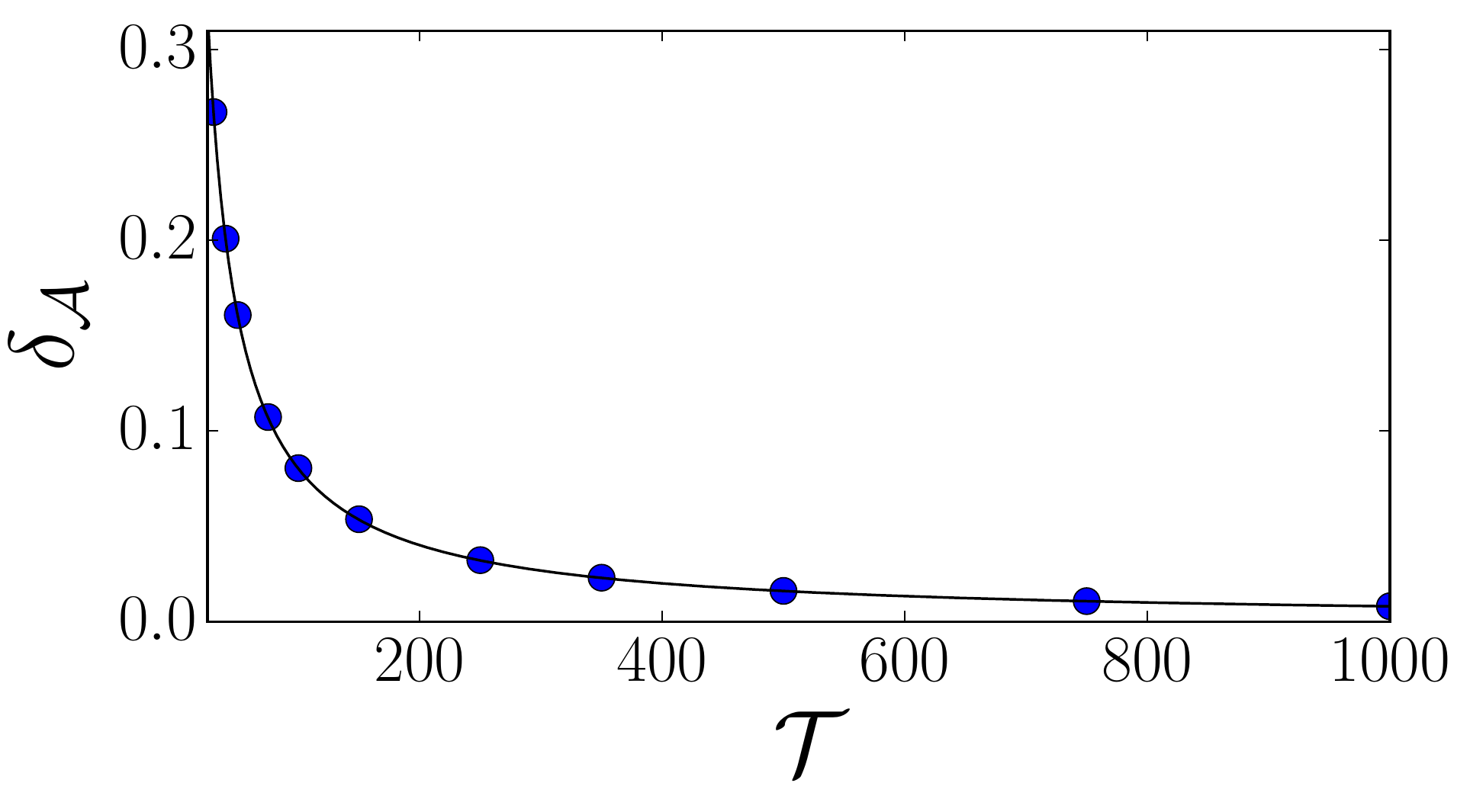}
\caption{A numerical estimation of the deviations of the cycle areas $\delta_{\cal A}$ in Eq.$\eqref{area_C}$ for the 2-levels Otto engine in Fig.\ref{otto_plot}. We see an approach $\delta_{\cal A} \propto 1/\T$ to the quasi-static limit with a proportionality constant $c=8.02$, extracted from our fitting datas.}\label{area_otto}
\end{figure}

Finally, we show that within our framework one is able to engineer different cycles. For instance, an Otto engine is obtained by replacing $\Omega$ in Eq.$\eqref{Omega_carnot}$ with
\be\label{Omega_otto}
\Omega(t)=\begin{cases} \Omega_1 \\ 2\Omega_1 -\Omega_2 +\frac{4(\Omega_2-\Omega_1)}{\T} t \\ \Omega_2 \\ -3\Omega_1 + 4\Omega_2 +\frac{4(\Omega_1-\Omega_2)}{\T} t\end{cases}
\ee
and the temperature in Eq.$\eqref{temperature}$ with
\be\label{temp_otto}
T(t)=\begin{cases} \frac{4(T_b-T_a)}{\T}t+ T_a \\ - \\ \frac{4(T_d-T_c)}{\T}t+3T_c-2T_d \\ - \end{cases},
\ee
where the first and the third isothermal strokes are substituted by thermal isochoric strokes. One can easily show that the quasi-static reversibility conditions now read
\be\label{otto_QS}
\frac{\Omega_2}{\Omega_1}= \frac{T({\bf c})}{T({\bf b})}=\frac{T({\bf d})}{T({\bf a})}
\ee 
and are satisfied in Eq.$\eqref{temp_otto}$ if $T_c=(\Omega_2/\Omega_1) T_b$, $T_d=(\Omega_2/\Omega_1) T_a$. In the Fig.\ref{floq_gamma_otto} and \ref{otto_plot} we show the Floquet parameters and the operation of the Otto engine respectively, obtained by a numerical solution of Eq.$\eqref{Y_eq}$. The deviations $\delta_{\cal A}$ in the cycle area from the quasi-static limit value are plotted in Fig.\ref{area_otto} and decay as a power law $\delta_{\cal A}\sim {\cal O}(1/\T)$  for large $\T$.

\section{Summary and conclusions}\label{conclusions}
In this article, we have presented an algebraic method for the solution of time-dependent Lindblad master equations for open quantum systems of finite Hilbert space dimension. The approach relies on the existence of a superoperator basis 
 for the Liouvillean generator and on the fact that the elements of such basis form a closed algebra (see appendix \ref{algebra_superop}). 
It follows that the dynamical map admits a canonical parametrisation as  product of exponentials of the basis elements with coordinates that are complex functions. This allows us to cast the initial problem into finding  solutions of first-order differential equations for the coordinates of the map which is, at least numerically, readily solvable in most cases.\\
We also addressed the problem of a periodic driving. We first considered an alternative formulation of our method that may be viewed as a generalisation of a rotating frame transformation, where a map $W_t$ connects the initial time-dependent Liouvillean $\Li_t$ to a time-independent one $\widetilde{\Li}$. The time evolution is then completely determined by first solving the time-independent problem and then by mapping the solution back to the original frame. 
Afterwards, we showed that requiring time periodicity in the map $W_t=W_{t+\T}$, it may be possible to write the time-dependent master equation in the form of a Floquet theory for the open system. We remark that the markovianity of the stroboscopic evolution is not guaranteed \cite{Schnell18} and must be checked through e.g. the tests proposed in the Refs. \cite{Wolf08,Cubitt12}. \\
We proposed some applications concerning the case of a single qubit. In particular, we studied the Lindblad evolution of the spin subject to finite temperature amplitude damping and dephasing noise, formally in the general case and explicitly for two specific choices of the parameters (see Sec.\ref{counter_oscillating_case} and Sec.\ref{incoherent_driving}).\\
In the last section, we applied our framework to the analysis of finite-time quantum heat-engines \cite{Alicki-79,Kosloff-Levy,Alicki2015,Rezek2006,Abah2014,Klaers2017a,Ronagel2014,Correa2014,Jaramillo2016,Samuelsson2017,DelCampo2014,Abah2016,Abah2017,Camati2016,Elouard2017,Cottet2017,Masuyama2017,Manzano2017b,Micadei2017,Manzano2017,Scopa-18}, since all the physical informations about the cyclic evolution (such as the limit cycle state or the time scale of the transient regime) can be deduced from the knowledge of the Floquet-Liouvillean $\Li_F$.
For concreteness, we engineered a quantum heat-engine operating with a working fluid made of a two-levels system. From the general discussion, we explicitly solved for a Carnot cycle (see Fig.\ref{carnot_plot}) and for an Otto cycle (see Fig.\ref{otto_plot}), both operating in a finite-time. The deviations from the quasi-static operation have been estimated through the deviation in the cycle area $\delta_{\cal A}$ which we propose as a quantifier of the non-equilibrium behaviour of the device.\\
We mention that the main limitation of this framework is related to the manipulation of the algebraic structures and so to the derivation of the first-order differential equations. However, we truly believe that it is possible to improve our achievements using more sophisticated numerical implementations. This would allow, for instance, to extend our results to spin chains and to investigate for them transport phenomena arising from the driving.
We leave this improvements to future developments. As a further perspective, we would like to investigate rigorously the possible connection between the existence of periodic solutions defining $W_t$ and the markovianity conditions for the Floquet dynamical map, recently introduced in the Ref.\cite{Schnell18}.
%

%
%
%
%
%
%
%
%
%
%
%
%
\indent\\
{\it Acknowledgements.} The authors acknowledge the USP-Cofecub project number Uc Ph $167$-$17$. AH would like to acknowledge the LPCT of Nancy for the warm hospitality during his internship period. 

\appendix

\section{General construction of $\mathfrak{su}(n)$ generators}\label{app_complete_set}
In the following, we recall a general procedure to build the generators of $\mathfrak{su}(n)$ Lie algebras \cite{Georgi-book}.  Consider the ensemble of $n\times n$ matrices $P^{ik}$:
\be
\left(P^{ik}\right)_{\mu\nu}= \delta_{i\mu}\delta_{k\nu} \quad i,k,\mu,\nu=1,\dots,n
\ee
in which all the elements are zero except one. By definition, we have 
\be\label{ortho_PP}
\tr(P^{ik}\, P^{jl})=\delta_{il}\delta_{kj}\,.
\ee
For $i<k$, we define the $n(n-1)/2$ real and the $n(n-1)/2$ imaginary combinations:
\be
{\cal K}^{ik}= \frac{1}{\sqrt{2}}\left(P^{ik}+P^{ki}\right)\,, \quad {\cal J}^{ik}= -\frac{i}{\sqrt{2}}\left(P^{ik}-P^{ki}\right)
\ee
while using $P^{ii}$ we construct the $n-1$ traceless diagonal matrices
\be
{\cal M}^q= \frac{1}{\sqrt{q(q+1)}}\sum_{k=1}^{q} \left( P^{kk}- q\,P^{\,q+1,q+1}\right)
\ee
with $q=1,\dots,n-1$. From Eq.$\eqref{ortho_PP}$, we see that the $n^2-1$ matrices $ {\cal K}^{ik}, {\cal J}^{ik}, {\cal M}^q$ are orthonormal and thus they form a basis of $\mathfrak{su}(n)$.

\section{Algebra of superoperators}\label{algebra_superop}
Here, we report the commutation relations of the superoperators $\eqref{sup_basis1}$-$\eqref{sup_basis2}$ for a generic $n$-levels open quantum system ${\cal S}$. 
First, it is easy to show that the set of unitary operators $\{{\cal H}_j\}_{j=1}^{n^2-1}$ satisfies a $\mathfrak{su}(n)$ algebra:
\be
[{\cal H}_i, {\cal H}_j]= \sum_{k=1}^{n^2-1} \f_{ijk} {\cal H}_k
\ee
where $\f_{abc}\equiv -i \tr([F_a,F_b]\, F_c)$ is the standard antisymmetric tensor of $\mathfrak{su}(n)$. The commutator between the unitary $\{{\cal H}_j\}_{j=1}^{n^2-1}$ and the non-unitary set $\{ \D_{kl}\}_{k,l=1}^{n^2}$ reads instead
\be
[{\cal H}_j, {\cal D}_{kl}]=\sum_{s=1}^{n^2-1} ( \f_{jks} \, {\cal D}_{sl} + \f_{jls} \, {\cal D}_{ks} )
\ee
while non-unitary elements have a commutator
\begin{widetext}
\be\begin{split}
[\D_{kl}, \D_{qm}]=&\frac{1}{16} \sum_{s,p,r=1}^{n^2-1} \z_{lks} \, \z_{mqp} \, \f_{spr} \, {\cal H}_r +\frac{1}{4} \sum_{s,p=1}^{n^2-1} (\z_{mqs} ( \f_{ksp} \,\D_{pl} + \f_{slp} \,\D_{kp} ) + \z_{lks} ( \f_{sqp}\, \D_{pm} + \f_{msp}\, \D_{qp}))\\
& +\frac{1}{4} \sum_{s,p=1}^{n^2-1} (\z_{qks} \,\z_{lmp} - \z_{kqs}\,\z_{mlp}) \D_{sp} +\frac{1}{2n} \sum_{s=1}^{n^2-1} (\delta_{qk} \, \f_{mls} - \delta_{lm} \, \f_{kqs}) {\cal H}_s\,,
\end{split}\ee
\end{widetext}
where $\text{z}_{abc}\equiv (\f_{abc} -i \dd_{abc})$ and $\dd_{abc}\equiv\tr(\{ F_a,F_b\} F_c)$ is the symmetric tensor of $\mathfrak{su}(n)$.

\section{Coherence vector formalism}\label{coherence}

\begin{figure}
\includegraphics[scale=0.55]{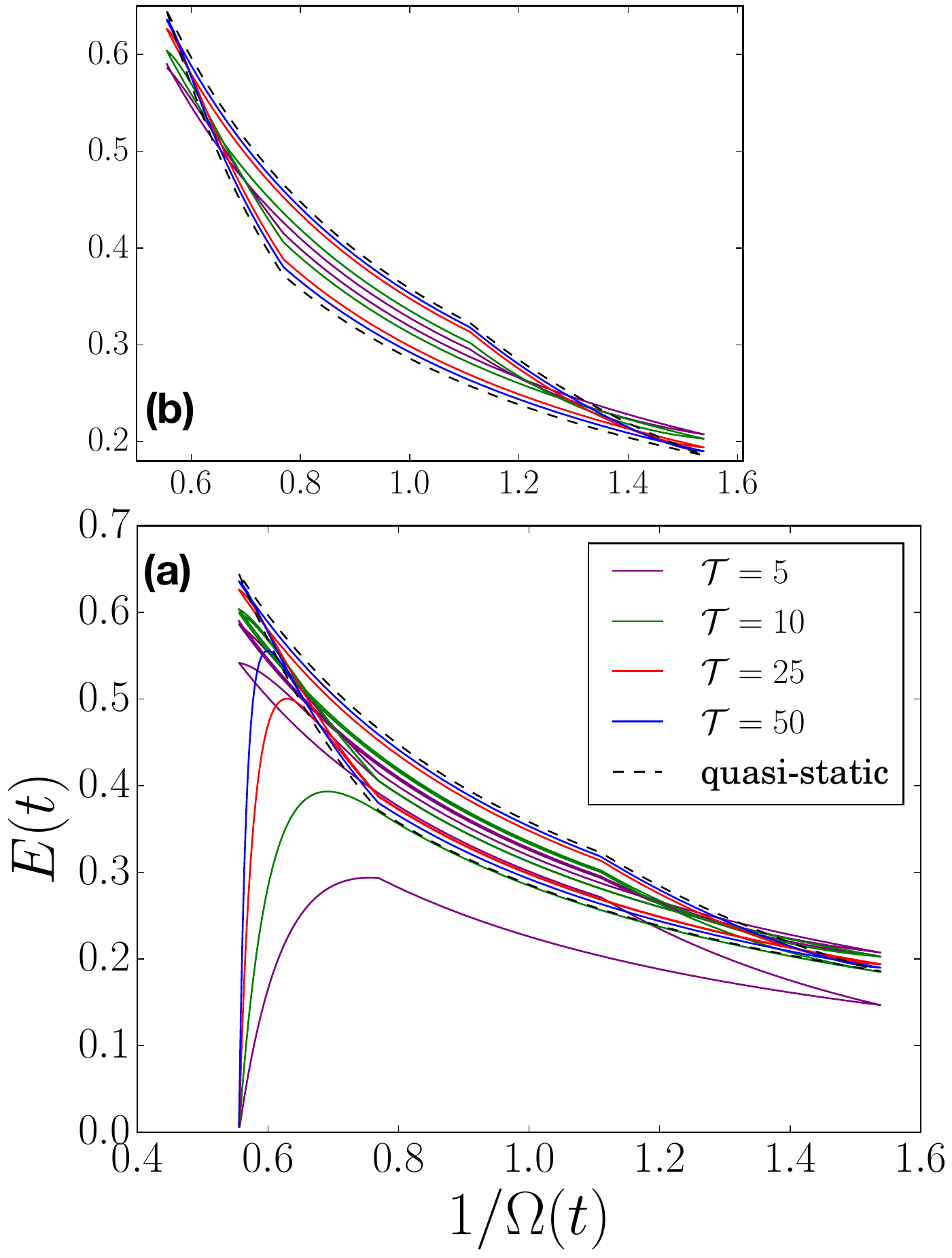}.
\caption{Operation of a finite-time Carnot engine obtained by a numerical solution of Eqs.$\eqref{coh_eq}$ with initial state $\rho_{0}=\Id_2/2$. The plots show the internal energy $E(t)$ as function of the compression $1/\Omega(t)$. In the figures we select $\Omega$ in Eq.$\eqref{Omega_carnot}$ with $\Omega_a=1.8$, $\Omega_b=1.3$ and the thermal bath in Eq.$\eqref{thermal_bath_gamma}$ with $T\hot=1.0$, $T\cold=0.5$. {\bf(a)} We show the solution in a time window $t\in [0,3\T]$ for different values of the driving period $\T$. We can see that at large enough values of $\T$, the cycle matches its quasi-static limit. {\bf(b)} Inset showing the time evolution after $\T/\Gamma$ where the dynamics becomes cyclic. The results are then in perfect agreement with those in Fig.\ref{carnot_plot}.
}\label{coh_carnot_engine}
\end{figure}

We report a matrix formulation of the markovian master Eq.$\eqref{Lind_eq}$ known under the name of coherence vector formalism \cite{Alicki-Lendi}. The basic idea of the method is to encode all the informations about the reduced density matrix $\rho_t$ in a $(n^2-1)$-vector:
\be
\vec{v}(t)=\left( v_1(t),\dots,v_{n^2-1}(t)\right) \in \mathbb{R}^{n^2-1}
\ee
where 
\be
v_k(t)=\tr(\rho_t\, F_k)\,, \quad k=1,\dots, n^2-1\,,
\ee
that we call coherence vector. In terms of this vector, the Lindblad equation becomes a sort of Bloch equation in $\mathbb{R}^{n^2-1}$:
\be\label{bloch_eq}
\frac{d}{dt}\vec{v}(t)= (Q(t)  + R(t) )\, \vec{v}(t) + \vec{k}(t)\,.
\ee
Here the matrix $Q(t)$, defined as
\be\label{Q_matrix}
(Q(t))_{sk}=\sum_{q=1}^{n^2-1} \f_{qks} \, h_q(t) \in \mathbb{R}\,,
\ee
is responsible for the unitary evolution. Notice that the unitarity is reflected in the conservation of the vector norm
\be
\frac{d}{dt} \norm{\vec{v}(t)}^2=\vec{v}(t)^{\text{T}}( Q(t)^{\text{T}}+ Q(t)) \vec{v}(t)=0\,,
\ee
following from the skew symmetry $Q(t)=-Q(t)^{\text{T}}$ of the matrix in Eq.$\eqref{Q_matrix}$. The purely dissipative contribution of the Lindblad Eq.$\eqref{Lind_eq_2}$  is instead represented through the matrix 
\be\label{R_matrix}
(R(t))_{qs}= -\frac{1}{4} \sum_{i,k,l=1}^{n^2-1} \gamma_{ik}(t) (\text{z}^*_{ilq} \f_{kls} + \z_{klq} \f_{ils})
\ee
and by the vector 
\be\label{k_vec}
k_s(t)=\frac{i}{n}\sum_{i,k=1}^{n^2-1} \gamma_{ik}(t) \, \f_{iks}\,,
\ee
both real quantities, as shown in \cite{Alicki-Lendi}. In the Eqs. $\eqref{Q_matrix}$, $\eqref{R_matrix}$ and $\eqref{k_vec}$ above, we adopted the same notations for the structure constants $\f_{abc}$, $\dd_{abc}$  of the appendix \ref{algebra_superop}.\\
From a geometrical point of view, the time evolution in Eq.$\eqref{bloch_eq}$  is a kind of rotation in $\mathbb{R}^{n^2-1}$ with a varying vector length: whereas rotations are generated by Hamiltonian and non-Hamiltonian terms, the variations of the vector length are due to dissipative effects only. \\
In the case $n=2$ and in particular, for the Lindblad evolution in Eq.$\eqref{Lindblad_2lvl}$, the Bloch Eq.$\eqref{bloch_eq}$ leads to the set of equations:
\begin{subequations}\label{coh_eq}
\be
\frac{d}{dt} a(t)-\Omega(t)\,b(t) +(\alpha(t)+2\Gamma_3(t))a(t)=0\,;
\ee
\be
\frac{d}{dt} b(t)+\Omega(t)\,a(t) +(\alpha(t)+2\Gamma_3(t))b(t)=0\,;
\ee
\be
\frac{d}{dt} \rho_{11}(t)+ 2\alpha(t) \rho_{11}(t) -(\alpha(t)+\beta(t))=0\,,
\ee
\end{subequations}
for the coherence vector components
\be
\vec{v}(t)=\sqrt{2} \bV a(t) \\ b(t) \\ \rho_{11}(t)-1/2 \eV
\ee
where we set $\rho_{12}(t)\equiv a(t)-ib(t)$. For concreteness, we consider the two-levels Carnot engine described in Sec.\ref{carnot_2lvl} and we investigate its time evolution by solving the Eqs.$\eqref{coh_eq}$. In  Fig.\ref{coh_carnot_engine}, we show the results for the operation cycle: we can clearly see the presence of an initial transient of duration ${\cal T}/\Gamma$ (compare with Eq.$\eqref{transient}$) after which the system behaves cyclically.

\section{Magnus high-frequency expansion}\label{high_freq}
For large enough driving frequencies $\omega\equiv2\pi/\T\gg 1$, the auxiliary frame method of Sec.\ref{rotating_frame} can be used to derive approximate results in agreement with those obtained by means of Magnus expansions \cite{Dai-16,Dai-17}.  Indeed, without loss of generality, we write the Liouvillean as
\be
\Li_t= \overline{\Li} + V_t
\ee
where $\overline{\Li}\equiv\T^{-1}\int_0^{\T} dt \, \Li_t$ and $V_t\equiv\Li_t -\overline{\Li}$. For future convenience, we also introduce the Fourier series of $V_t$
\be
V_t= \sum_{k\neq0} e^{ik\omega t}\,  \hat{V}_k\,.
\ee
At this point, we define the map $W_t$ to the auxiliary frame $\widetilde{\cal S}$ as 
\be
W_t= e^{\Phi_t}
\ee
with $\Phi_t=\Phi_{t+\T}$ a linear combination of the superoperators in $\eqref{sup_basis1}$-$\eqref{sup_basis2}$ and we consider the following expansions in powers of $\omega^{-1}$:
\be\label{HFexp}
\Phi_t= \sum_{a=1}^{\infty} \omega^{-a} \, \Phi^{(a)}_t\,, \quad \widetilde{\Li}=\sum_{a=0}^{\infty} \omega^{-a} \widetilde{\Li}^{(a)}.
\ee
According to Eq.$\eqref{aux_Li}$, we then obtain
\be\begin{split}\label{aux_Li_exp}
\widetilde{\Li}=&e^{\Phi_t}\, \Li_t \, e^{-\Phi_t} + (\frac{d}{dt} e^{\Phi_t})\, e^{-\Phi_t}\\
=& \Li_t + \frac{d}{dt} \Phi_t +[\Phi_t,\Li_t] + \frac{1}{2}[\Phi_t, [\Phi_t,\Li_t]] + \dots
\end{split}\ee
which can be solved in high-frequency expansion $\eqref{HFexp}$ imposing the time independence of $\widetilde{\Li}$ order by order. The results read at the second order:
\begin{widetext}
\be
\widetilde{\Li}=\overline{\Li} + \frac{1}{\omega}( \sum_{k\neq 0} \frac{i}{2k} [\hat{V}_k,\hat{V}_{-k}]) -\frac{1}{\omega^2}(\sum_{k\neq0} \frac{1}{2k^2} [[\hat{V}_k,\overline{\Li}],\hat{V}_{-k}] + \sum_{k,q\neq 0}\frac{1}{3kq} [\hat{V}_k,[\hat{V}_{q},\hat{V}_{-k-q}]]) +{\cal O}(\omega^{-3})
\ee
and
\be
\Phi_t= \frac{i}{\omega}\sum_{k\neq0} \frac{e^{ik\omega t}}{k} \hat{V}_k -\frac{1}{\omega^2}\sum_{k\neq 0}\sum_{q\neq -k} \frac{e^{i(k+q)\omega t}}{2k(k+q)} [\hat{V}_k,\hat{V}_q] +{\cal O}(\omega^{-3})\,.
\ee
\end{widetext}
For what concerns the Floquet generator $\Li_F$ in Eq.$\eqref{Floq_Li}$, we have
\be\begin{split}
\Li_F&=e^{-\Phi_{t_0}} \, \widetilde{\Li} \, e^{\Phi_{t_0}}\\
&=\widetilde{\Li} +[\Phi_{t_0},\widetilde{\Li}] +\frac{1}{2}[\Phi_{t_0},[\Phi_{t_0}, \widetilde{\Li}]]+\dots
\end{split}\ee
and similarly for the micromotion in Eq.$\eqref{kick}$ 
\be\begin{split}
\mathscr{K}_{t,t_0}&=e^{-\Phi_t}\, e^{\Phi_{t_0}}\equiv e^{K_{t,t_0}}\\
&=\exp( -\Phi_t + \Phi_{t_0} -\frac{1}{2}[\Phi_t,\Phi_{t_0}]+\dots)\,.
\end{split}\ee
In high-frequency expansion $\eqref{HFexp}$, the latter lead to the expressions:
\begin{widetext}
\be\begin{split}
\Li_F&= \overline{\Li} +\frac{i}{\omega}\sum_{k\neq0}\frac{\frac{1}{2}[\hat{V}_k,\hat{V}_{-k}]-[\hat{V}_k,\overline{\Li}]}{k}-\frac{1}{\omega^2}\sum_{k\neq0} \frac{\frac{1}{2}[[\hat{V}_k,\overline{\Li}],\hat{V}_{-k}]-[[\hat{V}_k,\overline{\Li}],\overline{\Li}]}{k}\\
&+\frac{1}{\omega^2}\sum_{k,q\neq 0} \frac{\frac{1}{2}[\hat{V}_k,[\hat{V}_q,\overline{\Li}]]-\frac{1}{2}[\hat{V}_k,[\hat{V}_q,\hat{V}_{-q}]]+\frac{1}{3}[\hat{V}_k,[\hat{V}_q,\hat{V}_{-k-q}]]-\frac{1}{2}[[\hat{V}_k,\hat{V}_{q-k}],\overline{\Li}]}{kq} +{\cal O}(\omega^{-3})\,,
\end{split}\ee
\be\begin{split}\label{kick_HF}
K_{t,t_0}= &\frac{i}{\omega}\sum_{k\neq 0} \frac{(e^{ik\omega t_0}-e^{ik\omega t})}{k} \hat{V}_k -\frac{1}{\omega^2}\sum_{k\neq 0} \frac{(e^{ik\omega t_0}-e^{ik\omega t})}{k^2} [\hat{V}_k,\overline{\Li}] \\
&-\frac{1}{\omega^2}\sum_{k,q\neq 0} \frac{(e^{iq\omega t_0}-e^{iq\omega t})[\hat{V}_k,\hat{V}_{q-k}]- e^{ik\omega t}\,e^{iq\omega t_0}[V_k,V_q]}{2kq} + {\cal O}(\omega^{-3})
\end{split}\ee
\end{widetext}
which extend the results in \cite{Rahav03,Goldman-PRX-14} to the dissipative case.\\
 \begin{figure}[ht]
 \includegraphics[scale=0.4]{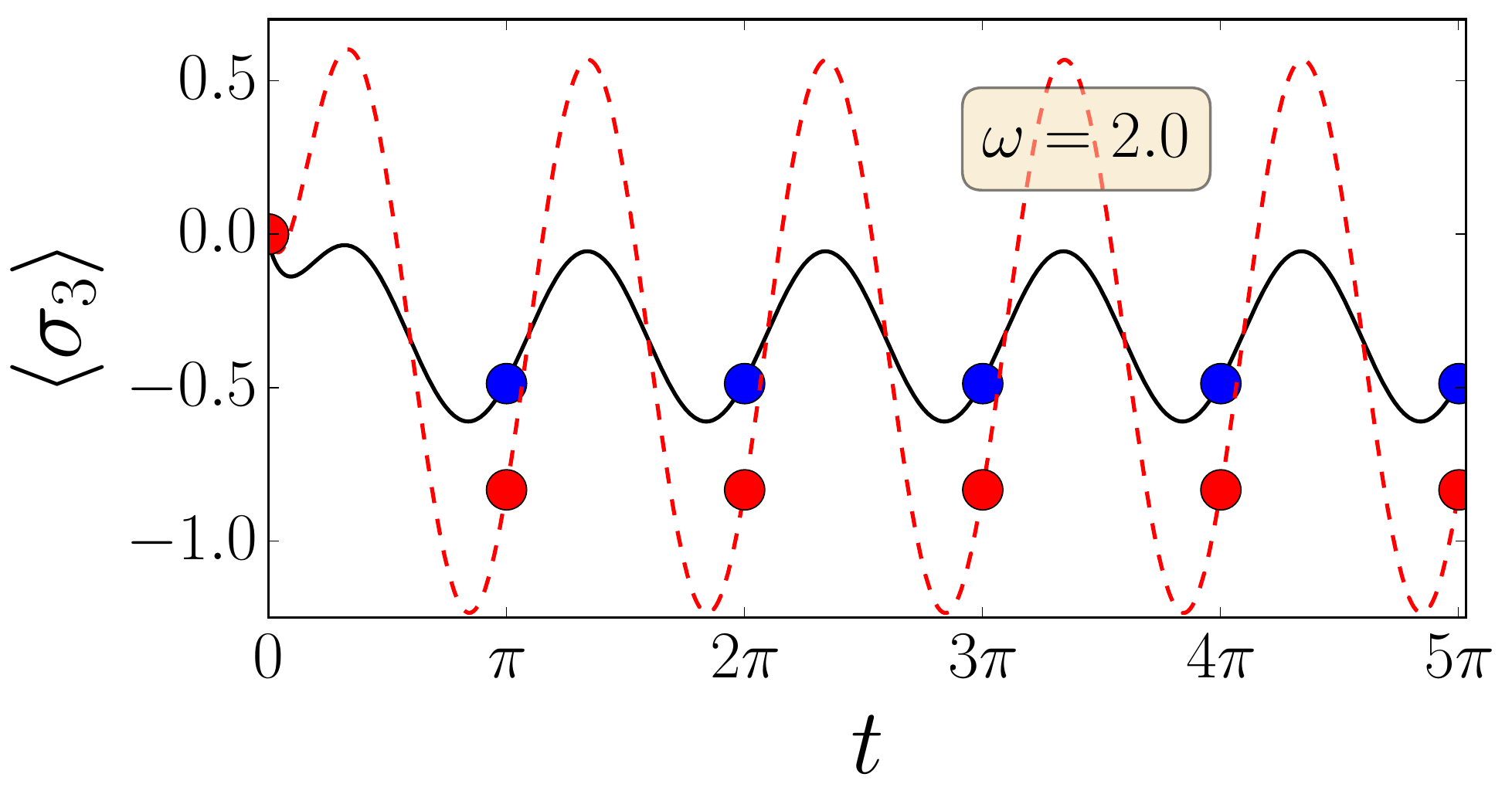}  \includegraphics[scale=0.4]{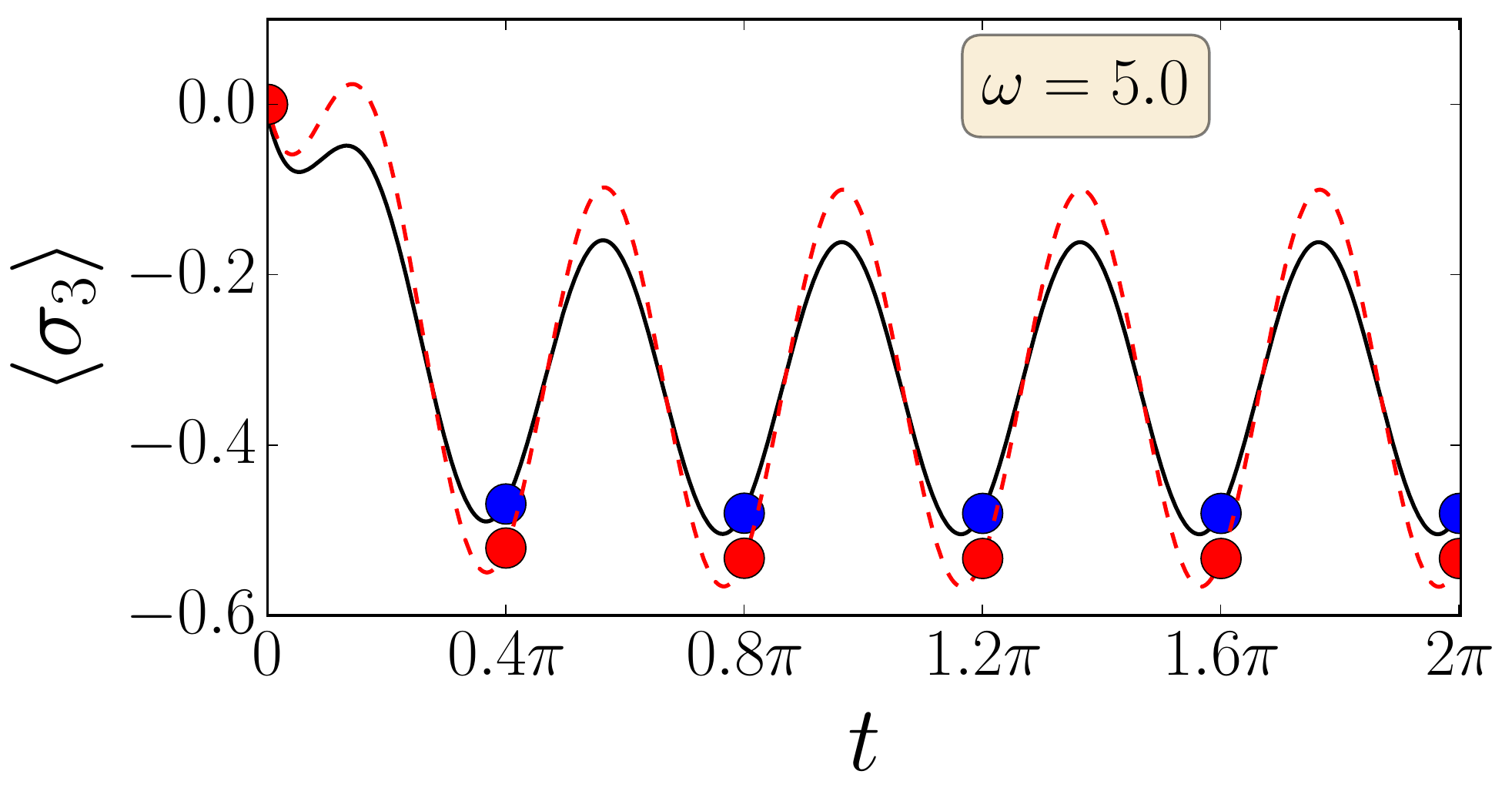} \includegraphics[scale=0.4]{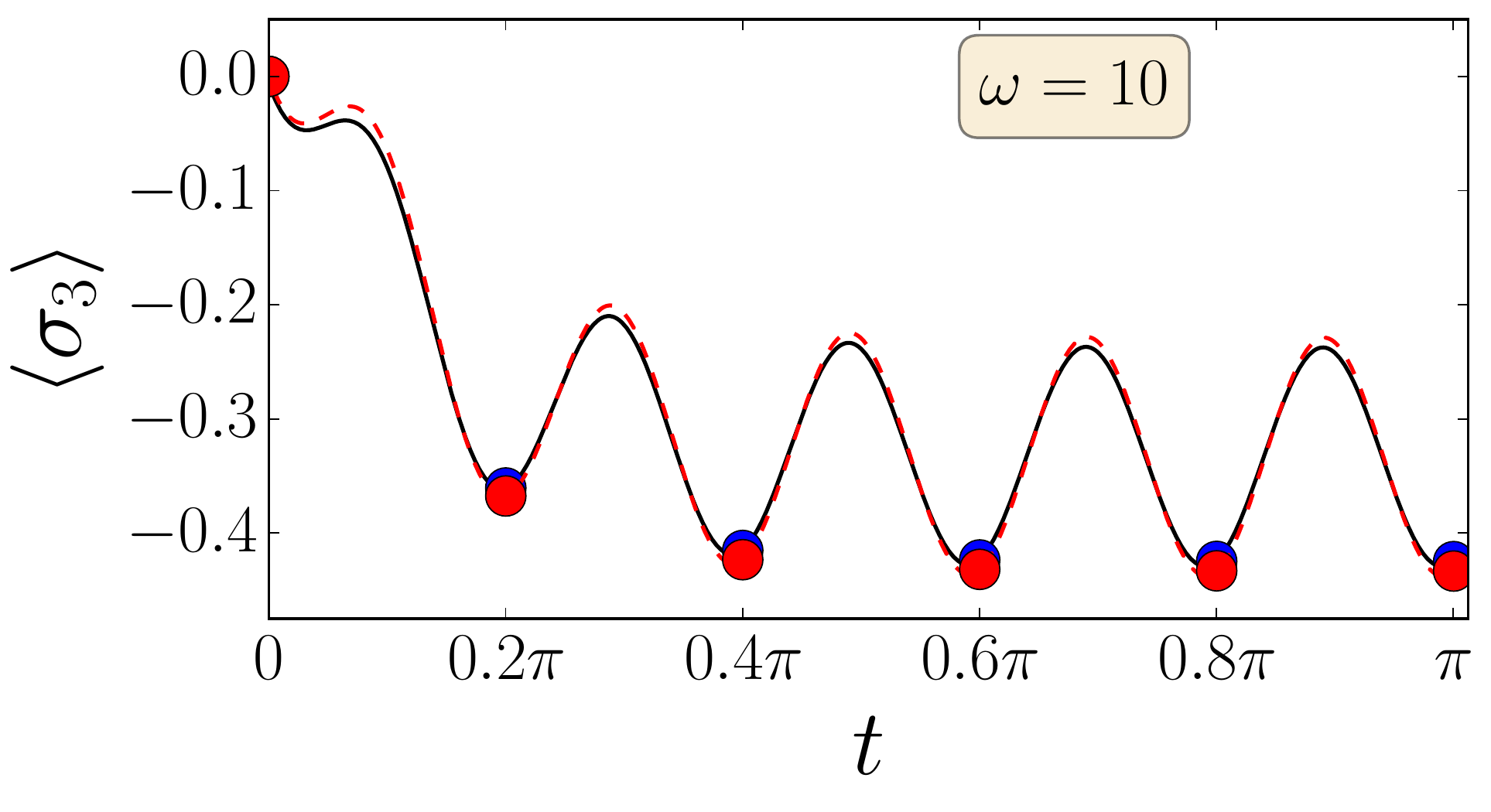}
 \caption{Time evolution of the spin magnetisation $\braket{\sigma_3}=\tr(\rho_t\,\sigma_3)$ for the single qubit in Eq.$\eqref{Lindblad_2lvl}$ with counter-oscillating polarisers $\Gamma_{1}=2.0+ 0.5 \sin(\omega t)$, $\Gamma_-=3.0- 0.5 \sin(\omega t)$ and with $\Gamma_3=0$, $\Omega=\sqrt{2}(1-\cos(\omega t))$. The numerical exact time evolution ({\it full line}) is compared with the approximate time evolution ({\it dashed line}) obtained in high-frequency perturbation theory at the second order for three values of the driving frequency $\omega=2.0,\,5.0,\,10$. The dots show the stroboscopic evolution (in {\it blue}: exact, in {\it red}: approximate).
 The system is prepared at time $t=0$ in a state $\rho_0=\Id_2/2$.}\label{compare_HF}
 \end{figure} 
 
We apply the high-frequency expansion to the example of a single qubit subject to counter oscillating polarisers (see Eq.$\eqref{counter_pol}$) discussed in Sec.\ref{counter_oscillating_case}. It follows that the set of Floquet parameters is 
\be\label{shift_HF}
\delta\Gamma(t_0)= \frac{A\Gamma}{\omega} + {\cal O}(\omega^{-3})\,,
\ee
 $\Omega^F= \overline{\Omega}+{\cal O}(\omega^{-3})$ and $\Gamma_3^F= \overline{\Gamma}_3+{\cal O}(\omega^{-3})$ in agreement with the exact solution in Eq.$\eqref{counter_shift}$, while for the micromotion one obtains
 \be\begin{split}\label{counter_kick_HF}
 K_{t,0}=&\frac{A}{\omega}(1-\cos(\omega t))(\D\p-\D\m) \\
 &+\frac{1}{\omega^2}(A\Gamma (\D\p-\D\m) + \Delta\, {\cal H}_3) \sin(\omega t) +{\cal O}(\omega^{-3})\,,
 \end{split}\ee
 setting $t_0=0$ and $\Omega=\overline{\Omega}+\Delta\,\cos(\omega t)$. The time evolution of the spin magnetisation obtained in high-frequency approximation is shown in Fig.\ref{compare_HF} for different values of the driving frequency. As expected, the larger $\omega$, the better the agreement with the exact solution.


\section{Basic concepts of quantum thermodynamics}\label{quantum_therm}
In this appendix we review some elementary concepts of quantum thermodynamics useful for the investigation of quantum heat-engines.
We may define, roughly speaking, a quantum heat-engine as a quantum device capable to convert the thermal energy extracted from a hot source into a power output.
To do so, an abstract quantum heat-engine needs:
\begin{itemize}
\item an open quantum system ${\cal S}$ that is used as {\it working fluid},
\item a {\it hot} and a {\it cold} thermal baths ${\cal R}$ through which the engine can extract work,
\item Some periodic external fields that determine its {\it operation cycle}.
 \end{itemize}
The operation of the engine can be monitored by looking at the time variations of thermodynamic observables $X_t \in \text{End}(\mathscr{H}_{\cal S})$. In the weak-coupling limit, the dynamics of $X_t$  is generated by the Lindblad equation of Heisenberg form:
\be\label{dual_Lind}
\frac{d}{dt} X_t = \frac{\de}{\de t} X_t + i[H_t, X_t] + D^{\star}_t(X_t)\,,
\ee
where the dual Lindblad dissipator $D^{\star}_t$ reads 
\be
D^{\star}_t(\bullet) =\sum_{k,l=1}^{n^2-1} \gamma_{kl}(t) (F_l^{\dagger} \, \bullet \, F_k -\frac{1}{2}\{F_l^{\dagger}F_k, \bullet \,\})\,.
\ee
Defining the internal energy of the system as $E(t)=\braket{H_t}$, with $\braket{\, \bullet\, }\equiv\tr(\rho_t \,\bullet\,)$, we have from Eq.$\eqref{dual_Lind}$
\be\label{first_law}
\frac{d}{dt} E(t)= \left\langle\frac{\de}{\de t} H_t\right\rangle + \left\langle D^{\star}_t(H_t)\right\rangle
\ee
which is nothing but the quantum analogue of the first law of thermodynamics in a differential form \cite{Alicki-79,Geva-92}. Indeed, we may interpret the two terms in the r.h.s. of Eq.$\eqref{first_law}$
\be
{\cal P}(t)=\left\langle\frac{\de}{\de t} H_t\right\rangle\,, \quad \delta{\cal Q}(t)=\left\langle D^{\star}_t(H_t)\right\rangle\,,
\ee
as the power output ${\cal P}$ and the heat flow $\delta{\cal Q}$ of the engine. From the latter, one can eventually define the {\it efficiency} $\eta$ of the engine as the ratio between the rate of work generated and the rate of thermal energy extracted from the bath i.e., $\eta={\cal P}/{\delta{\cal Q}}$.\\

 \bibliography{draftbib}
\end{document}